\documentclass{article}
\usepackage{graphicx} 
\usepackage{amsfonts} 
\usepackage{mathptmx}       
\usepackage{helvet}         
\usepackage{courier}        
\usepackage{amsmath,bm}
\usepackage{amsthm}
\usepackage{amssymb}
\usepackage{subfigure}
\usepackage{blindtext}
\usepackage{hyperref}
\usepackage{apacite}
\usepackage{natbib}
\usepackage{type1cm}        
\usepackage[toc,page]{appendix}
\usepackage{makeidx}         
\usepackage{graphicx}        
\usepackage{multicol}        
\usepackage[bottom]{footmisc}
\usepackage{colortbl,booktabs}
\usepackage{tablefootnote}
\usepackage{enumitem} 
\usepackage{hyperref}

\usepackage{algorithm}
\usepackage{algpseudocode}

\newcommand{\indep}{\perp \!\!\! \perp}
\newcolumntype{M}[1]{>{\centering\arraybackslash}m{#1}}

\newtheorem{prop}{Proposition}

\title{Bayesian Nonparametric Models for Multiple Raters: a General Statistical Framework}
\author{Giuseppe Mignemi, Ioanna Manolopoulou}
\date{}

\begin{document}

\maketitle

\begin{abstract}
Rating procedure is crucial in many applied fields (e.g., educational, clinical, emergency). It implies that a rater (e.g., teacher, doctor) rates a subject (e.g., student, doctor) on a rating scale. Given raters’ variability, several statistical methods have been proposed for assessing and improving the quality of ratings. The analysis and the estimate of the Intraclass Correlation Coefficient (ICC) are major concerns in such cases. As evidenced by the literature, ICC might differ across different subgroups of raters and might be affected by contextual factors and subject heterogeneity. Model estimation in the presence of heterogeneity has been one of the recent challenges in this research line. Consequently, several methods have been proposed to address this issue under a parametric multilevel modelling framework, in which strong distributional assumptions are made. We propose a more flexible model under the Bayesian nonparametric (BNP) framework, in which most of those assumptions are relaxed. By eliciting hierarchical discrete nonparametric priors, the model accommodates clusters among raters and subjects, naturally accounts for heterogeneity, and improves estimates' accuracy. We propose a general BNP heteroscedastic framework to analyze continuous and coarse rating data and possible latent differences among subjects and raters. The estimated densities are used to make inferences about the rating process and the quality of the ratings. By exploiting a stick-breaking representation of the Dirichlet Process, a general class of ICC indices might be derived for these models. Our method allows us to independently identify latent similarities between subjects and raters and can be applied in \textit{precise education} to improve personalized teaching programs or interventions. Theoretical results about the ICC are provided together with computational strategies. Simulations and a real-world application are presented, and possible future directions are discussed.
\end{abstract}


 \clearpage

\section{Introduction}\label{intro}
Rating procedure is crucial in several applied scientific fields, such as educational assessment \citep{Tasha_2023, Chin_2020}, psychological and medical diagnoses \citep{DLIMA2024, Krolikowska2023,li2022}, emergency rescue \citep{albrecht2024clinical,lo2021inter} or grant review process \citep{sattler2015grant, Cao_2010}. It implies that an observer, commonly called a rater (e.g., teacher, doctor), assesses some subject attribute or latent ability (e.g., student proficiency, patient severity) on a rating scale. 
Raters' variability might pose reliability concerns and uncertainty about the quality of ratings \citep{Bartos24, Mignemi2024,tenHove2021}. Several statistical methods have been proposed to address these issues, they aim to assess or improve the accuracy of ratings \citep{Martinkova2023, Casabianca_2015, Nelson2015, Gwet_Li, McGraw1996}.  
Multilevel modelling serves as a natural statistical framework for rating data since subjects are either nested within raters or crossed with them \citep{tenHove2021}. These models (e.g., one-way or two-way ANOVA, hierarchical linear or generalized linear models) decompose the total variance of observed ratings according to different sources of variability, i.e. subjects and raters \citep[see][chap. 4, for an overview]{martinkovabook}.
The observed rating is commonly broken down into different effects, for instance, the effect of the subject (i.e., true score, latent ability; \citealt{lord_novick68}), the effect of the rater (i.e., rater's systematic bias) and a residual part \citep{McGraw1996,shrout1979intraclass}.
This allows us to jointly estimate the subject true score and the reliability of ratings, which is generally referred to as the proportion of total variance due to the subjects' variability \citep{McGraw1996, Joreskog_74}. \\
Several methods have been proposed to analyze rating data under the Item Response Theory (IRT) framework, such as the Generalized Many Facet Rasch Models \citep[GMFRMs;][]{Uto_24, Uto_2020,linacre1989many}, the Hierarchical Raters Models \citep[HRMs;][]{Molenaar2021, nieto2019, decarlo2011, Patz2002} or the Generalized Hierarchical Raters Models \citep[GHRMs;][]{Muckle2009}. These models jointly estimate the subject's latent ability, rater effects (e.g., systematic bias and reliability), and item features (i.e., difficulty, discrimination). They typically rely on the assumption that subjects' latent abilities are independent and identically distributed (i.i.d.) from a normal distribution. 
Other recent research lines concentrate on modelling and estimation issues in the presence of subjects' and raters' heterogeneity \citep{Martinkova2023,ten2022updated,sattler2015grant,mutz2012heterogeneity}. These works model systematic differences among subjects or raters are to allow more accurate estimates and detailed information about the rating procedure. Individual subjects' or raters' characteristics may affect rating reliability, so that more flexible models result in separate reliability estimates \citep{Martinkova2023}. Recent models have been proposed to address this issue under a parametric multilevel modelling framework \citep{Martinkova2023, Erosheva2021, Martinkova2018,mutz2012heterogeneity} in which heterogeneity is addressed as a covariate-dependent difference among subjects and subject- and rater-specific effects are assumed to be i.i.d from a normal distribution. \\
The normality assumption made under all the aforementioned models might be unrealistic under a highly heterogeneous scenario in which possible clusters among subjects or raters might be reasonably expected and the conditional density of the respective effects might be multimodal \citep{Paganin2023, Dunson2010, Verbeke1996}. Such patterns have emerged from real data, showing that both the conditional densities of subjects' latent ability \citep[e.g.,][]{Uto_24} and raters' systematic bias \citep[e.g.,][]{Muckle2009} might be multimodal and the normality assumption violated. In these cases, the data exhibit two levels of heterogeneity. The first, known as \textit{individual} heterogeneity, captures the differences between individuals; the second, referred to as \textit{population} heterogeneity, pertains to the differences between clusters. Although parametric mixture models might represent a suitable solution, the number of mixture components needs to be fixed. Models with different numbers of components have to be fitted and model selection techniques are required to identify the optimal number of clusters \citep{Bartholomew2011}. \\
\subsection{Our Contributions}
Our proposal aims to overcome these restrictions under a Bayesian nonparametric (BNP) model, which naturally accommodates subgroups among students and raters and allows less restrictive distributional assumptions on the respective effects \citep{Ghosal_van_der_Vaart_2017, Hjort_Holmes_Müller_Walker_2010, Ferguson_1973}. Bayesian nonparametric inference has led to new developments and advances during the last decades in psychometrics \citep{Roy2024, Paganin2023,cremaschi2021bayesian, Wang2020, tang2017, SanMartin2011, Dunson2010, Karabatsos2009}, but to the best of our knowledge, it has never been applied to rating data modelling.   
We provide a flexible statistical framework for rating models in which latent heterogeneity among subjects and raters is captured with the stochastic clustering induced by the Dirichlet Process Mixture (DPM) placed over their respective effects. Modelling subjects' and raters' effect parameters as an infinite mixture of some distribution family (e.g., Normal, Gamma) enables the model to account for possible multimodality without specifying the number of mixture components \citep{DeIorio_2023, Dunson2010}. Although previous works have raised questions about the identifiability of the parameters in BNP IRT models \cite{SanMartin2011}, theoretical results by \citet{Pan_2024} have recently shown that BNP IRT models (e.g., 1PL) are identifiable.\\
Under the general case of a two-way design \citep{McGraw1996}, we specify a measurement model for the subject latent ability (e.g., student proficiency) in which the rater's systematic bias (i.e., severity) and reliability are consistently estimated. This makes our method more relevant for subject scoring purposes than the other Bayesian nonparametric models proposed for the analysis of rating data \citep{Kottas2018bayesian, savitsky2014bayesian, kottas2005nonparametric}. Our proposal may be suitable both for balanced (i.e. when all raters score each subject; \citealt{Nelson2015, Nelson2010}) and unbalanced designs (i.e. when a subset of raters scores each subject; \citealt{ten2022updated, Martinkova2023}).  Furthermore, we propose a Semiparametric model as a nested version of the BNP in which raters' effects are i.i.d. from a unimodal distribution. Very small rater sample sizes may not reasonably be considered representative of the overall rater population, making the semiparametric specification a potentially more suitable choice.\\
The advantages of the proposed method are manyfold. First, it relies on more relaxed distributional assumptions for the subjects' and raters' effects, allowing for density estimation using mixtures \citep{Ghosal1999, Escobar_95} and preventing model misspecification issues \citep{antonelli2016, Walker2007}. As recently argued by \citet{tang2017}, Bayesian nonparametric priors might be helpful in assessing the appropriateness of common parametric assumptions for psychometrics models and represent a solution under their violation \citep{Antoniak_1974, Ferguson_1973}. Second, it naturally enables independent clustering of subjects and raters, bringing more detailed information about their latent differences \citep{Mignemi2024, DeIorio_2023}. This allows the joint analysis of \textit{individual} and \textit{population} heterogeneity of both subjects and raters. This aspect might be beneficial in the context of \textit{precise education} \citep{coates_2025_precision, Cook_2018_precision}, where information about individual and cluster differences might be used for implementing more personalized educational programs or interventions \citep{hart_2016_precision, henderson_2020_precise}. Third, exploiting a stick-breaking representation of the Dirichlet Process \citep{Ghosal_van_der_Vaart_2017, Ishwaran2001}, a general class of ICC indices might be derived, and different indices might be computed according to distinct clusters of subjects or raters. Fourth, it is readily extended to account for coarse or ordinal ratings \citep{Lockwood2018, Coarse_2015a}. Fifth, the general hierarchical formulation of our model allows comparisons with other methods and further extensions under unifying modelling frameworks (e.g., generalized linear latent and mixed model, GLLAMM \citealt{rabe2016generalized}). This facilitates a straightforward communication between different statistical fields and a wider application of the BNP method. \\
Model parameters are learned through full posterior sampling. Since most of the parameters in the model have conjugate prior distributions, full conditional Gibbs sampling is possible for most of the parameters \citep{Ishwaran2001}. Nonetheless, few parameters do not have conjugate priors and a derivatives matching technique is involved to approximate the full conditional \citep{Miller2019}. 

\subsection{Outline of the Paper}
The outline of the paper is as follows: we present the general framework and introduce the model in Sections ~\ref{Frame}-\ref{Main}, respectively; different approximate ICC indices are derived in Section ~\ref{ICC} and a reduced model for one-way designs is detailed in Section ~\ref{Reduced};  prior elicitation and posterior sampling are discussed and presented in Section ~\ref{Inference}; simulations and real-world applications are illustrated, respectively, in Section ~\ref{Simulation} and Section ~\ref{Applications}; the model extension for coarse ratings and is presented in Section \ref{Ord_ex}, along with some numerical results from real and generated data. Advantages and limitations of the proposal are discussed in Section ~\ref{Discussion}. Further Bayesian nonparametric extensions, proofs for ICCs indices, and additional plots are given in the Appendices. 
Additional results on balanced design in small sample sizes, technical details on out-of-sample predictive performance assessment and posterior computation for this class of models are presented in the Supplementary Materials. We provide an R package {\fontfamily{qcr}\selectfont
RatersBNP} to facilitate direct usage by researchers and practitioners of our method. Code and Supplementary Materials are available online through the link: \url{https://osf.io/3yx4j/?view_only=98c600198a6b4807878989765118f97e}.

\section{BNP Rating Model}
\subsection{General Framework}\label{Frame}  
Several model specifications have been proposed for different data structures and designs\citep{ten2022updated, Gwet_Li,shrout1979intraclass}. One-way designs are preferred when rater differences are
typically considered as noise \citep{Martinkova2023}, whereas two-way designs are usually involved if the rater's effect needs to be identified \citep{Mignemi2024, Casabianca_2015}. Balanced designs require each subject to be rated by all the raters, while in an unbalanced design each subject is only rated by a generally small subset of them \citep{tenHove2021}. Raters might be considered either fixed or random (i.e., drawn from the population) depending on the inference the researcher might be interested in \citep{koo2016}. \\  
The unbalanced two-way design with random raters is considered a general case to present our model. The reasons for this choice are both theoretical and practical. We aim to provide a comprehensive statistical framework for modelling the dependency of ratings on different categorical predictors (i.e., subjects' and raters' identities). This setting is a neat compromise between the one-way design, which implies only one categorical predictor (i.e. subject identity), and more complex dependency structures that involve more than two identities (i.e., several categorical predictors). Our proposal might be alternatively reduced or extended to be suitable for these different levels of complexity. The unbalanced design implies some sparsity in the co-occurrence between subjects and raters and each subject is rated only by a small subset of raters \citep{papaspiliopoulos2023, Omiros_19}, as a consequence each rater might score a different number of subjects. This makes the framework general and flexible, it might be seen as an extension of cross-classified models in which uncertainty is modelled also hierarchically. 
From a practical perspective, our choice is reasonable since many large studies and applications use unbalanced designs to distribute the workload across different raters \citep{ten2022updated}. \\

\subsection{Preliminaries on Bayesian Nonparametric Inference}\label{Preliminaries}
In this subsection, we briefly review some basic preliminaries on Bayesian nonparametric (BNP) inference providing here a very general framework which is detailed in Sections below (refer to \citealp{Ghosal_van_der_Vaart_2017} and \citealt{Hjort_Holmes_Müller_Walker_2010} for exhaustive treatments). \\
Suppose $Y_1,\dots, Y_n$, are observations (e.g., ratings), with each $Y_i$ taking values in a complete and separable metric space $\mathbb{Y}$. 
Let $\Pi$ denote a prior probability distribution on the set of all probability measures $\mathbf{P}_{\mathbb{Y}}$ such that:  
\begin{equation}\label{eq:0}
    Y_i | p  \overset{\mathrm{iid}}{\sim} p, \quad \;\; p \sim \Pi, 
\end{equation}
for $i=1,\dots,n$. Here $p$ is a random probability measure on $\mathbb{Y}$ and $\Pi$ is its probability distribution and might be interpreted as the prior distribution for Bayesian inference \citep{DeBlasi2015}. The inferential problem is called parametric when $\Pi$ degenerates on a finite-dimensional subspace of $\mathbf{P}_{\mathbb{Y}}$, and nonparametric when the support of $\Pi$ is infinite-dimensional \citep[chap. 3]{Hjort_Holmes_Müller_Walker_2010}. To the best of our knowledge, the vast majority of the contributions present in rating models literature  \citep{Bartos24, Martinkova2023,ten2022updated,tenHove2021, Martinkova2018, Zupanc2018, Casabianca_2015, Nelson2015, Nelson2010} are developed within a parametric framework making use of a prior that assigns probability one to a small subset of $\mathbf{P}_{\mathbb{Y}}$. Although \citet{Mignemi2024} recently proposed a Bayesian semi-parametric model for analyzing rating data. Even if they relax the normality assumption for the rater effect (i.e., the systematic bias), normality is still assumed for the subject true score distribution. This strong prior assumption is overcome through a BNP approach \citep{Ghosal_van_der_Vaart_2017} in the present work. 
\\

\paragraph{Dirichlet Processes.}
For the present proposal, we assume $\Pi$ to be a discrete nonparametric prior and correspond to a Dirichlet process (DP) which has been widely used in BNP psychometric research \citep{Paganin2023,cremaschi2021bayesian, Dunson2010, Karabatsos2009}. Given $\Pi = DP(\alpha P_0)$, $p$ is a random measure on $\mathbb{Y}$ following a DP with concentration parameter $\alpha>0$ and base measure $P_0$. This implies that for every finite measurable partition $\{B_1,\dots,B_k\}$ of $\mathbb{Y}$, the joint distribution $(p(B_1),\dots,p(B_k))$ follows a $k$-variate Dirichlet distribution with parameters $\alpha P_0(B_1),\dots,\alpha P_0(B_k)$: 
\begin{equation}\label{DP0}
    (p(B_1),\dots,p(B_k)) \sim Dir(\alpha P_0(B_1),\dots,\alpha P_0(B_k)). 
\end{equation}
The base measure $P_0$ is our \textit{prior guess} at $p$ as it is the prior expectation of the DP, i.e. $\mathbf{E}[p]=P_0$. The parameter $\alpha$ (also termed precision parameter) controls the concentration of the prior for $p$ about $P_0$. In the limit of $\alpha \rightarrow \infty$, the probability mass is spread out and $p$ gets closer to $P_0$; on the contrary, as $\alpha \rightarrow 0$, $p$ is less close to $P_0$ and concentrates at a point mass. \\

\paragraph{Dirichlet Process Mixtures.}Given the discrete nature of the DP, whenever $\mathbb{Y} = \mathbb{R}$ it is not a reasonable prior for the real-valued random variable $Y$. Nonetheless, it might be involved in density estimation through hierarchical mixture modelling \citep{Ghosal_van_der_Vaart_2017}. Let $f(\cdot;\Tilde{\theta})$ denote a probability density function for $\Tilde{\theta} \in \Theta \subseteq \mathbb{R}$, we modify (\ref{eq:0}) such that for $i=1,\dots,n$:
\begin{equation}\label{eq:0bis}
    Y_i | \Tilde{\theta}_i \overset{\mathrm{ind}}{\sim}  f(\cdot;\Tilde{\theta}_i), \quad 
    \Tilde{\theta}_i|p \overset{\mathrm{iid}}{\sim} p, \quad \;\; p \sim DP(\alpha P_0). 
\end{equation}
The realizations of the DP are almost surely (a.s.) discrete which implies a positive probability that $\Tilde{\theta}_i = \Tilde{\theta}_{i'}$, for $i \neq i'$. Indeed, a random sample $(\Tilde{\theta}_1,\dots, \Tilde{\theta}_n)$ from $p$ features $1\leq K_n \leq n$ different unique values $(\Tilde{\theta}^*_1,\dots, \Tilde{\theta}^*_{K_n})$ and leads to a random partition of $\{1,\dots, n\}$ into $K_n$ blocks such that $ \Tilde{\theta}_i \in (\Tilde{\theta}^*_1,\dots, \Tilde{\theta}^*_{K_n})$ for $i=1,\dots,n$. This naturally induces a mixture distribution for the observations $Y_1,\dots, Y_n$ with probability density: 
\begin{equation}\label{DPM_int}
   f(Y) = \int f(Y;\Tilde{\theta})p(d\Tilde{\theta}).
\end{equation}
To provide some intuition, by using a DP as a prior for an unknown mixture distribution we mix parametric families nonparametrically \citep{Gelman2014}. This model specification introduced by \citet{Lo84} and termed Dirichlet Process Mixture (DPM) provides a BNP framework to model rating data. 

\subsection{Proposed Model}\label{Main}
Consider a subject $i = 1, \dots, I$, whose attribute is independently scored by a random subset of raters $\mathcal{R}_i \subseteq \{1,\dots, J\}$ on a continuous rating scale. We assume that the observed rating $Y_{ij} \in \mathbb{R}$ depends independently on subject $i$ and rater $j \in \mathcal{R}_i$. The effect of the former is interpreted as $i$'s true score and is the rating procedure's focus. We let the residual part, that is the difference between the true and the observed score, depend on rater $j$'s effects, i.e. systematic bias and reliability.

\paragraph{Modelling Rating $Y_{ij}$.} 
We specify the following decomposition of rating $Y_{ij}$:
\begin{equation}\label{eq:1}
    Y_{ij} = \theta_i + \tau_j +\epsilon_{ij},  \quad i = 1,\dots,I; \;\; j \in \mathcal{R}_i. 
\end{equation}
Here $\theta_i$ captures the subject $i$'s latent "true" score and $ \tau_j +\epsilon_{ij}$ is the difference between the observed and the true score, representing the error of rater $j$. We assume these terms to be mutually independent.     

\begin{figure}
    \centering
   \includegraphics[scale=0.35]{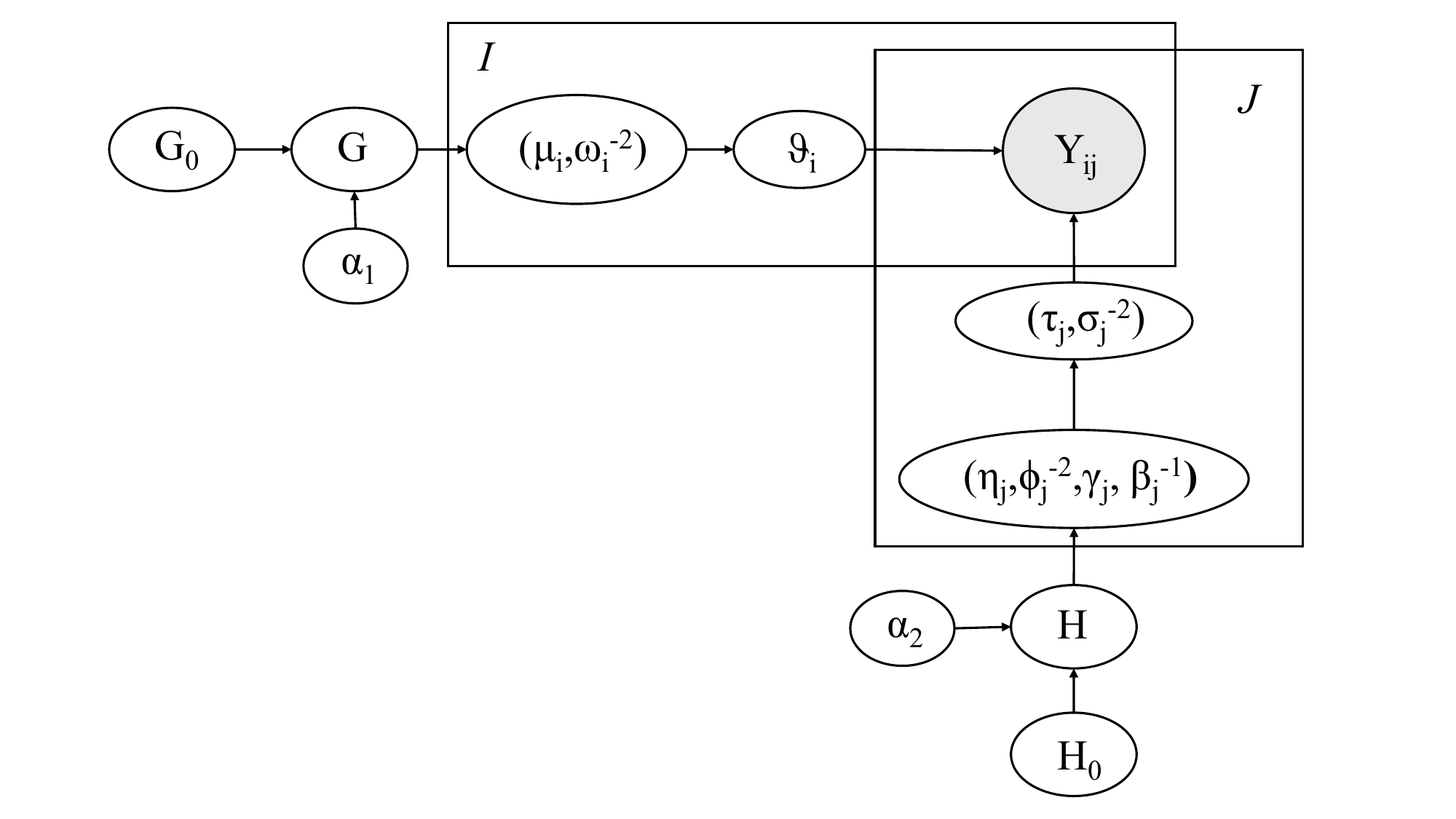}
    \caption{\small  Graphical representation of the dependencies implied by the model. The boxes indicate replicates, the four outer plates represent, respectively, subjects and raters, and the inner grey plate indicates the observed rating.}
    \label{Path}
\end{figure}

\paragraph{Modelling Subject's True Score.}\label{structural_1}
For each subject $i = 1,\dots,I$ we assume that the true score $\theta_i$ is independently distributed following a normal distribution with mean $\mu_i$ and variance $\omega^2_i$:
\begin{eqnarray}\label{eq:2}
     \theta_i|\mu_i,\omega^2_i & \overset{\mathrm{ind}}{\sim} & N(\mu_i,\omega^2_i).
\end{eqnarray}
Here $\mu_i$ is the mean of subject $i$'s true score, $\omega^2_i$ is its variability and we assume them to be independent. Conditional on the rater's error, higher values of $\theta_i$ imply higher levels of the subjects' attribute (e.g. higher student proficiency); on the contrary lower values indicate poor levels of their attribute (e.g. poor student proficiency).  \\
We specify a DP prior with precision parameter $\alpha_1$ and base measure $G_0$ for the pair $(\mu_i,\omega^2_i)$, $i=1,\dots,I$: 
\begin{equation}\label{DP1}
    (\mu_i,1/\omega^2_i)|G \overset{\mathrm{iid}}{\sim} G, \quad \; G \sim DP(\alpha_1G_0).
\end{equation}
We choose $G_0=N(\mu_0, S_0) \times Ga(w_0, w_0/W_0)$, where $\mu_0$ and $S_0$ are the mean and variance of the normal distribution and $w_0$ and $W_0$ are, respectively, the shape and the mean parameters of the gamma. 
We note that $G$ is a.s. discrete with a non-zero probability of ties, such that different subjects will share the same values of $(\mu_i,1/\omega^2_i)$ with a probability greater than zero, that is $P[(\mu_i,1/\omega^2_i)=(\mu_{i'},1/\omega^2_{i'})]>0$, for $i \neq i'$. This discreteness property naturally induces clustering across subjects and leads to a location-scale Dirichlet Process Mixture (DPM) prior for $\theta_i$. That is, this formulation can capture clusters of subject abilities. Figure \ref{Path} shows the hierarchical dependence of subjects' true scores.

\paragraph{Modelling Rater's Bias and Reliability.}\label{structural_2}
For each rater $j=1,\dots, J$, who scores a subset of subjects $\mathcal{S}_j \subseteq \{1,\dots,I\}: j \in \mathcal{R}_i$, the difference between the observed rating $Y_{ij}$ and the subject's true score $\theta_i$, $i \in \mathcal{S}_j$, is decomposed into the rater effects $\tau_j$ and $\epsilon_{ij}$ (\ref{eq:1}), assuming $\tau_j \indep \epsilon_{ij}$. We model $\tau_j$ to be normally distributed with mean $\eta_j$ and variance $\omega^2_j$: 
\begin{equation}\label{eq:3a}
    \tau_j|\eta_j,\phi^2_j \overset{\mathrm{ind}}{\sim} N(\eta_j,\phi^2_j), \quad j=1,\dots,J.
\end{equation}
Here $\eta_j$ and $\phi^2_j$ are the mean and the variance of the rater $j$'s effect $\tau_j$. It captures $j$'s specific systematic bias, i.e. the mean difference between the observed rating $Y_{ij}$ and the subject's true score $\theta_i$, $i \in \mathcal{S}_j$. Given two raters such that $\tau_j<\tau_{j'}$, 
$j$ is said to be more strict and expected to give systematically smaller ratings than $j$ on average.\\
The residual term $\epsilon_{ij}$ is assumed to be i.i.d. for $i \in \mathcal{S}_j$ following a normal distribution with zero mean and variance $\sigma^2_j$. We let this parameter vary across raters and assume $1/\sigma^2_j$ follows a gamma distribution with shape and rate parameters $\gamma_j,\gamma_j/\beta_j$, respectively: 
\begin{eqnarray}\label{eq:3b}
     \epsilon_{ij}|\sigma^2_j & \overset{\mathrm{iid}}{\sim} & N(0, \sigma^2_j), \quad i \in \mathcal{S}_j, \\ 
     1/\sigma^2_j|\gamma_j,\beta_j  & \overset{\mathrm{ind}}{\sim} & Ga(\gamma_j,\gamma_j/\beta_j), \quad j=1,\dots,J. \label{eq:3b_bis}
\end{eqnarray}
Under this parametrization, $1/\sigma^2_j$ is the rater $j$'s specific reliability with mean $\beta_j$ and $\gamma_j$ is the shape parameter. We prefer this parametrization for interpretability purposes, which implies a simpler notation below. Conditional on subjects' true score $\theta_i$, $i \in \mathcal{S}_j$, larger values of $\sigma^2_j$ imply more variability across the ratings given by $j$ and might be interpreted as a poorly consistent rating behaviour. On the contrary, smaller values of $\sigma^2_j$ indicate less variability and higher consistency for $j$ across subjects. \\
We specify a DP prior with concentration parameter $\alpha_2$ and base measure $H_0$ for the four-dimensional vector $(\eta_j,1/\phi^2_j,\gamma_j,1/\beta_j)$,  $j=1,\dots,J$:
\begin{equation}\label{DP2}
    (\eta_j,1/\phi^2_j,\gamma_j,1/\beta_j)|H \overset{\mathrm{iid}}{\sim} H, \quad \; H \sim DP(\alpha_2H_0).
\end{equation}
We assume mutual independence for the elements of the vector and choose $H_0=N(\eta_0, D_0) \times Ga(a_0, a_0/A_0) \times Ga(b_0, b_0/B_0) \times Ga(m_0, m_0/M_0)$, where $\eta_0$ and $D_0$  are mean and scale parameters, respectively; $a_0, b_0, m_0$ are shape parameters and $A_0, B_0, M_0$ are mean parameters. This formulation induces a DPM prior for raters' bias and reliability $\tau_j$ and $1/\sigma^2_j$. Figure \ref{Path} gives a graphical representation of the model. 
The independence assumption might be relaxed by employing a suitable multivariate base measure accounting for possible dependencies among the four elements of the vector. However, this implies a more complex specification, which is beyond the purpose of this work. Further constraints on raters' systematic bias $\tau$ are needed for identifiability purposes which are discussed in Section \ref{SC-DPM_sec}, after presenting the stick-breaking representation.

\subsection{Stick-breaking Representation}
The random probability measures $G$ and $H$ are assigned discrete priors, as a consequence they might be represented as a weighted sum of point masses: 
\begin{eqnarray}\label{sum}
   G &=& \sum_{n \geq1} \pi_{1n} \delta_{\xi_n}, \\
   H &=& \sum_{k\geq1} \pi_{2k} \delta_{\zeta_k}, \label{sum2}
\end{eqnarray}
where the weights $\{\pi_{1n}\}_{n=1}^{\infty}$ and $\{\pi_{2k}\}_{k=1}^{\infty}$ take values on the infinite probability simplex and $\delta_{x}(\cdot)$ stands for the Dirac measure and denotes a point mass at $x$. 
Note that, we index the components of the infinite mixture (\ref{sum}) corresponding to the subjects with $n=1,\dots,\infty$, whereas $k=1,\dots,\infty$ is used for that corresponding to the raters (\ref{sum2}).
The random vectors $\xi_n=(\mu_n,\omega^2_n)$, $n=1,\dots,\infty$ are i.i.d. from the base measure $G_0$, $\zeta_k=(\eta_k,\phi^2_k,\gamma_k,\beta_k)$, $k=1,\dots,\infty$ are i.i.d. from the base measure $H_0$, and both vectors are assumed to be independent of the corresponding weights. This makes clear why the expectations of the $DPs$ are $G_0$ and $H_0$, respectively and are said to be our \textit{prior guess} at $G$ and $H$ (see Section ~\ref{Preliminaries}). \\ 
This discreteness property of the $DP$ allows us to define $G$ and $H$ through the stick-breaking representation introduced by \citet{Sethuraman}:
\begin{equation}\label{StickBreaking1}
     G = \sum_{n \geq1} \pi_{1n} \delta_{\xi_n}, \quad 
  \pi_{1n}     = V_{1n} \prod_{l<n}(1-V_{1l}),   \quad 
  V_{1n}    \overset{\mathrm{iid}}{\sim}    Beta(1,\alpha_1),     \quad 
     \xi_n  \overset{\mathrm{iid}}{\sim} G_0,  
\end{equation}
 and 
 \begin{equation}\label{StickBreaking2}
     H = \sum_{k \geq1} \pi_{2k} \delta_{\zeta_k}, \quad 
  \pi_{2k}     = V_{2k} \prod_{l<k}(1-V_{2l}),   \quad 
  V_{2k}       \overset{\mathrm{iid}}{\sim} Beta(1,\alpha_2),     \quad 
     \zeta_k  \overset{\mathrm{iid}}{\sim} H_0. 
\end{equation}
This construction of the $DP$ implies that, for each subject $i=1,\dots, I$, $(\mu_i, \omega^2_i)=\xi_n$ with probability $ \pi_{1n} = V_{1n} \prod_{l<n}(1-V_{1l})$. Equivalently, for each rater $j=1,\dots,J$, the probability that $(\eta_j,\phi^2_j,\gamma_j,\beta_j)=\zeta_k$ is given by $\pi_{2k} = V_{2k} \prod_{l<k}(1-V_{2l})$. \\

\paragraph{Moments of student latent true score $\theta_i$.}
The mean and the variance of the subject's true score $\theta_i$, $i=1,\dots, I$, under a $DP(\alpha_1G_0)$ prior are: 
\begin{equation}\label{Moments_1}
 \mathbf{E}[\theta_i|G] = \mu_{G} =\sum_{n\geq 1} \pi_{1n} \mu_n, \quad \quad 
 \mathbf{Var}[\theta_i|G] = \omega^2_{G} =\sum_{n\geq 1} \pi_{1n} (\mu_n^2 + \omega_n^2) - \mu^2_{G},  
\end{equation}
where $\mu_n$ and $\omega_n^2$ are the mean and the variance of $\theta_i$ for the $n$-th component of the mixture. Here $\mu_{G}$ is the weighted average across components and captures the mean true score across subjects. The parameter $ \omega^2_{G}$ is the conditional variance of the infinite mixture and indicates the variability of true scores across subjects. 

\paragraph{Moments of raters'  bias $\tau_j$.}
The mean and the variance of the rater's bias $\tau_j$, $j=1,\dots, J$, under a $DP(\alpha_2H_0)$ prior are: 
\begin{equation}\label{Moments_2}
 \mathbf{E}[\tau_j|H] = \eta_{H} =\sum_{k \geq 1} \pi_{2k} \eta_k, \quad \quad 
 \mathbf{Var}[\tau_j|H] = \phi^2_{H} =\sum_{k \geq 1} \pi_{2k} (\eta_k^2 + \phi_k^2) - \eta^2_{H},  
\end{equation}
where $\eta_k$ and $\phi_k^2$ are the mean and the variance of $\tau_j$ for the $k$-th component of the mixture. Here $\eta_{H}$ and $\phi^2_{H}$ capture the mean and the variance of the systematic bias within the general population of raters.

\paragraph{Moments of raters' reliability $1/\sigma^2_j$.}
Raters' residual mean is fixed to zero by the model (\ref{eq:3b}), that is $\mathbf{E}[\epsilon]=0$; mean and variance of raters reliability $1/\sigma^2_j$ under a $DP(\alpha_2H_0)$ prior are: 
\begin{equation}\label{Moments_3}
 \mathbf{E}[1/\sigma^2_j|H] = \beta_{H} =\sum_{k \geq 1} \pi_{2k} \beta_k \quad \quad 
 \mathbf{Var}[1/\sigma^2_j|H] = \psi^2_{H} =\sum_{k \geq 1} \pi_{2k} (\beta_k^2 + \psi_k) - \beta^2_{H},   
\end{equation}
where $\beta_{H}$ captures raters' weighted average reliability and $\psi^2_{H}$ indicates the total reliability variance across them. Here $\beta_k$ and $\psi_k=\beta_k^2/\gamma_k$ are, respectively, the mean and the variance of $1/\sigma^2_j$ for the $k$-th component of the mixture.\\
Note that we model the independent rater's features, i.e. bias and reliability, by placing the same $DP(\alpha_2H_0)$ prior. In other terms, $\tau_j$ and $1/\sigma_j$ are two independent elements of the same vector drawn from $H$.  \\

\paragraph{Finite stick-breaking approximation.}\label{finite_sb}
The recursive generation defined in (\ref{StickBreaking1}) and (\ref{StickBreaking2}) implies a decreasing stochastic order of the weights $\{\pi_{1n}\}_{n=1}^{\infty}$ and $\{\pi_{2k}\}_{k=1}^{\infty}$ as the indices $n$ and $k$ grow. Considering the expectations $\mathbf{E}[V_{1n}]=1/(1+\alpha_1)$ and $\mathbf{E}[V_{2k}]=1/(1+\alpha_2)$ it is clear that the rates of decreasing depend on the concentration parameters $\alpha_1$ and $\alpha_2$, respectively. Values of these parameters close to zero imply a mass concentration on the first couple of atoms, with the remaining atoms being assigned small probabilities; which is consistent with the general formulation of the $DP$ discussed in Section \ref{Preliminaries}. Given this property of the weights, in practical applications the infinite sequences (\ref{sum}) and (\ref{sum2}), are truncated at enough large values of $R \in \mathbb{N}$: 
\begin{equation}\label{Trunc1}
     G = \sum_{n =1}^{R} \pi_{1n} \delta_{\xi_n}, \quad \quad  H = \sum_{k =1}^{R} \pi_{2k} \delta_{\zeta_k}.
\end{equation}
We use this finite stick-breaking approximation proposed by \citet{Ishwaran2001} to let $V_{1R}=V_{2R}=1$, and discard the terms $R+1,\dots,\infty$, for $G$ and $H$. \\
The moment formulas (\ref{Moments_1}), (\ref{Moments_2}) and (\ref{Moments_3}) are readily modified accordingly to the truncation and computed as finite mixture moments. 

\paragraph{Nested versions.} Semiparametric nested versions of the BNP model might be specified in which alternatively $G$ or $H$ are degenerate on a single component and $R=1$ for one of them in the finite approximation. That is, subjects or raters are all clustered together. For instance, for very small values of $J$ (i.e., raters' sample size), raters might not be reasonably considered a representative sample of their population and limited information is available for drawing inference about it. Under these scenarios, raters' effects might be assumed to be i.i.d. from a normal distribution.

\subsection{Semi-Centered DPM} \label{SC-DPM_sec}
Hierarchical models (e.g., GLMM, Linear Latent Factor models), might suffer from identifiability issues, and constraints on the latent variable distributions are needed for consistently identify and interpret model parameters \citep{Bartholomew2011, Dunson2010, gelman_hill_2006}. More specifically, under the linear random effects models a standard procedure to achieve model identifiability is to constrain the mean of the random effects to be zero \citep{Agresti2015}. We aim to consistently involve the same mean constraint for our proposal and allow straightforward and interpretable comparisons between the parametric and the nonparametric models. Similar to  \citet{Dunson2010}, we encompass a DPM-centered prior such that the expected value of the rater systematic bias is fixed to zero, $\mathbf{E}[\tau_j]=0$, for $j=1,\dots,J$.\\ 
Since the rating process focuses on the subjects' true scores, it might be more reasonable to centre the DPM for the raters' effects and let the model estimate the mean of the true scores $\mu_G$. Given that the mean of the raters' residual is fixed to zero in (\ref{eq:3b}), the mean raters' bias needs to be fixed. We adapt the centering procedure based on a parameter-expanded approach proposed by \citealt{Yang_2010} and \citealt{Dunson2010} to our proposal. We specify a semi-centered DPM (SC-DPM) involving an expansion in raters' systematic bias $\{\tau^*_j\}_1^J$, such that their mean $\eta^*_H=0$ a.s. The expanded-parameters (\ref{eq:3a}) can be expressed as:
\begin{equation}\label{SC-DPM}
   \tau^*_j = \tau_j - \eta_H,  \quad \quad 
      \tau_j|\eta_j,\phi^2_j \overset{\mathrm{ind}}{\sim} N(\eta_j,\phi^2_j), \quad \quad j=1,\dots, J,
\end{equation}
and the decomposition of rating $Y_{ij}$ (\ref{eq:1}) becomes: 
\begin{equation}\label{eq:1_exp}
     Y_{ij} = \theta_i + \tau^*_j +\epsilon_{ij}, \quad \quad i = 1,\dots,I; \;\; j \in \mathcal{R}_i. 
\end{equation}
Given the location transformation in (\ref{SC-DPM}) the expectation of the expanded parameters is zero: 
\begin{equation}\label{Exp_bias_new}
   \mathbf{E}[\tau^*_j|H] = 0.
\end{equation}
It is worth noting that the centering needs only to concern the location of the systematic bias and not its scale as it is in the centered-DPM introduced by \citet{Dunson2010}, which explains the term ``semi-centring" adopted here to avoid confusion. Accordingly, under the semiparametric specifications, the only location of the parametric distribution needs to be fixed; a zero mean normal distribution might be a suitable solution.

\section{BNP Intra-class Correlation Coefficient}\label{ICC}
Intra-class correlation coefficient (ICC) is widely used in applied statistics to quantify the degree of association between nested observations \citep{Agresti2015, Gelman2014} and to get relevant information about the level of heterogeneity across different groups \citep{Mulder_2019}.
Indeed, it is commonly applied in psychometrics to assess the consistency of ratings given by different raters to the same subject \citep{Martinkova2023,ten2022updated, Erosheva2021,tenHove2021, Nelson2015, Nelson2010}.
We provide a within-subject correlation structure (for any subject and a given raters pair) $ICC_{j,j'}$ based on the BNP model presented in Section \ref{Main}. This formulation relates to those proposed in psychometric literature regarding the $ICC_1$ (e.g., \citealt{Erosheva2021, DeBoeck_2008, fox_glas_2001, Bradlow_1999,shrout1979intraclass, Joreskog_74}), but doesn't rely on strong distributional assumptions and naturally accommodates for both subjects and raters sub-populations. We also propose a lower bound $ICC_A$ for the expected ICC which might be used for inference purposes about the general population of raters. An exact formula for the ICC suitable for the reduced one-way designs is proposed in Section \ref{Remarks_ICC_Reduced}. \\  
The paragraphs below provide preliminary information on computing the ICC under a parametric framework necessary to detail the BNP extension.

\paragraph{Parametric ICC.}
Under a parametric standard framework, i.e. equipping the parameters with finite-dimensional priors, the ICC is defined as the proportion of variance of the ratings due to the subjects' true score: 
\begin{equation}\label{ICC_p}
 ICC =  \frac{\omega_i^2}{\omega_i^2+\phi_j^2+\sigma_j^2} = \frac{\omega^2}{\omega^2+\phi^2+\sigma^2},
\end{equation}
assuming $\omega^2_i=\omega^2$, for $i=1,\dots, I$; $\phi^2_j=\phi^2$ and $\sigma^2_j=\sigma^2$, for $j=1,\dots, J$. Given two raters $j,j' \in \mathcal{R}_i, \;\; j \neq j'$ who rate the same subject $i$, the ICC is the correlation between the ratings $Y_{ij}$ and $Y_{ij'}$. Note that under this formulation $ICC \in [0,1]$, it can not capture any negative correlations. This index is also interpreted as the inter-rater reliability of a single rating and is also indicated by $IRR_1$ (see \citealt{Erosheva2021} for further details). \\
The homoscedastic assumption may be relaxed and raters' residual variance might be let to vary across raters according to (\ref{eq:3b}) and (\ref{eq:3b_bis}), given $\gamma_j=\gamma$ and $\beta_j=\beta$ for $j=1,\dots, J$.  

\begin{figure}
    \centering
    \subfigure{\includegraphics[scale= 0.38]{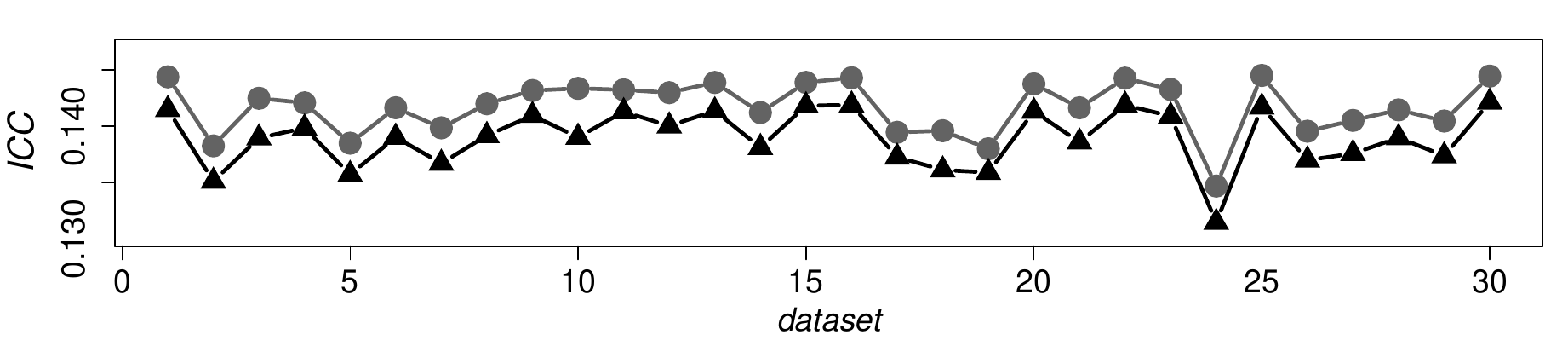}} 
    \subfigure{\includegraphics[scale= 0.38]{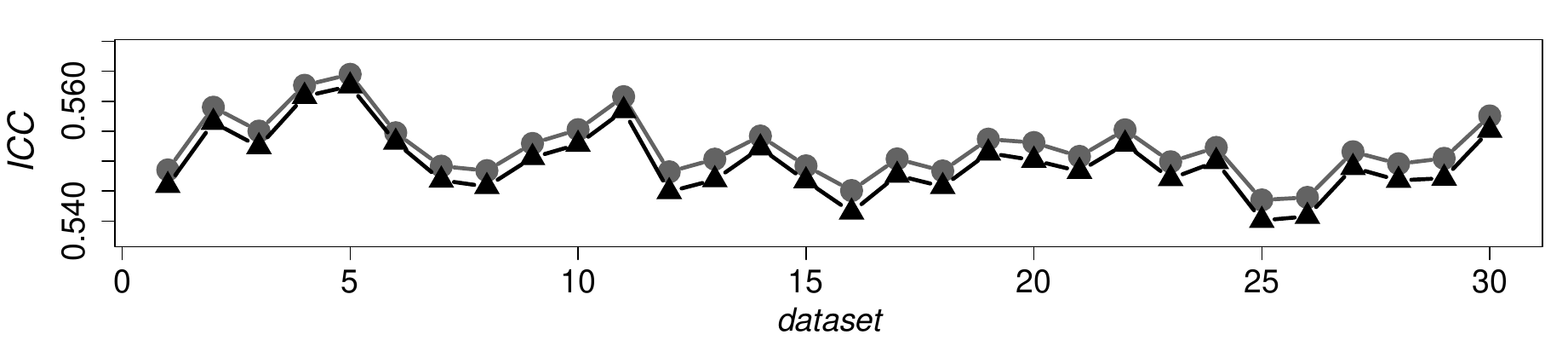}} 
     \subfigure{\includegraphics[scale=0.38]{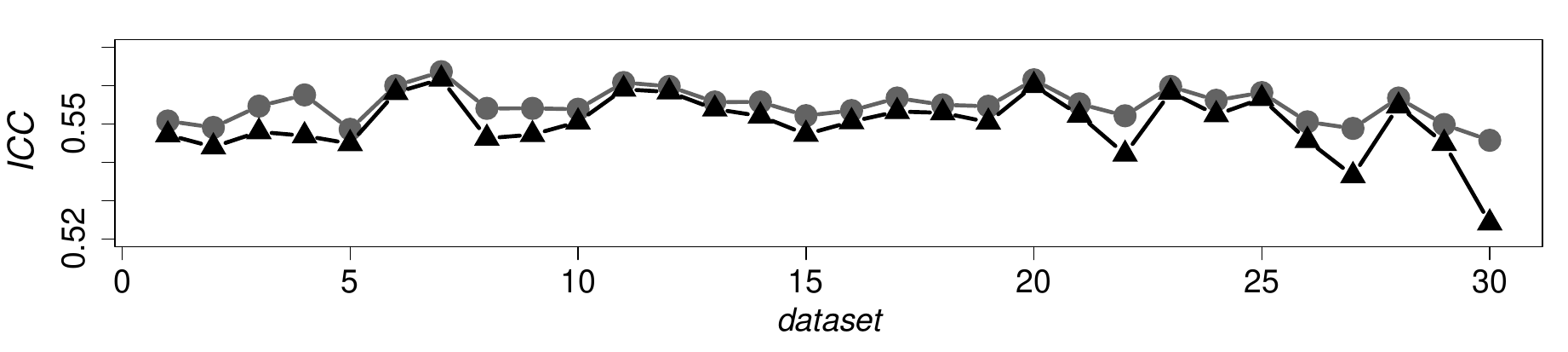}}
    \caption{\small Illustrative examples of empirical $ICC_A$ and $\mathbf{E}[ICC]$ across independent datasets and under different reliability scenarios. The grey balls indicate the mean pairwise $ICC$ between each rater and the others; the black triangles represent the computed $ICC_A$.}
    \label{ICCs}
\end{figure}

Given that $\sigma_j^2 \neq \sigma_{j'}^2$ for $j \neq j'$, it is possible to compute as many ICCs indices as possible pairs of raters, i.e. $J(J-1)/2$. 
In such cases the resulting $ICC_{j,j'}$ is the conditional correlation between the the ratings given to a random subject by raters $j$ and $j'$, given the other parameters: 
\begin{equation}\label{ICC_p_h}
 ICC_{j,j'} = \frac{\omega^2}{\sqrt{\omega^2+\phi^2+\sigma_j^2}\sqrt{\omega^2+\phi^2+\sigma_{j'}^2}}.
\end{equation}
A more general index accounting for all raters' residual variance might be more useful in applications. Despite the expected ICC, i.e. $\mathbf{E}[ICC|\omega^2,\phi^2]$, might represent a neat solution, it is not available in a close form and the posterior mean taken over the MCMC might be prohibitive in large scale assessments since there are $J(J-1)/2$ ICCs indices to compute for each iteration. An alternative index that might be readily computed is the ICC between two raters with average reliability. That is, we replace $\sigma_j^2$ with its expectation, i.e. $\mathbf{E}[\sigma^2]$: 
\begin{equation}\label{ICC_p_E}
 ICC_A = \frac{\omega^2}{\omega^2+\phi^2+\mathbf{E}[\sigma^2]}.
\end{equation}
It gives the correlation between the ratings given to the same random subject $i=1,\dots, I$ by two random raters $j,j' \in \mathcal{R}_i, \;\; j \neq j'$, satisfying $\sigma^2_j=\sigma^2_{j'}=\mathbf{E}[\sigma^2]$. That is the correlation between two ratings given to the same random student by two raters having an average reliability level. We note that they are different quantities: the expected pairwise ICC and the pairwise ICC between two mean reliable raters. Nonetheless, relying on a theoretical result that is given below, we can use the $ICC_A$ to have information about the other. \\
Given that the rater's reliability is assumed to follow a gamma distribution (\ref{eq:3b}), the inverse follows an inverse gamma distribution $\sigma^2_j|\gamma,\beta   \overset{\mathrm{ind}}{\sim}  IGa(\gamma,\gamma/\beta)$ for $j=1,\dots, J$, whose expected value is only defined for $\gamma>1$. In such cases we reparametrize (\ref{eq:3b}):
\begin{equation}\label{eq:3b_mod}
     1/\sigma^2_j|\gamma,\beta   \overset{\mathrm{iid}}{\sim}  Ga\left(1+\gamma,\frac{1+\gamma}{\beta}\right), \quad j=1,\dots,J.
\end{equation}
This specification ensures the expectation of raters' residual variance to be defined for any $\gamma>0$ and implies:
\begin{equation}\label{ExpVarRes}
    \mathbf{E}[\sigma^2_j|\gamma,\beta]=\Tilde{\sigma}= \frac{1+\gamma}{\beta \gamma}. 
\end{equation}
It is the mean raters' residual variance and its derivation is given in Supplementary Materials. The $ICC_A$ under the new parametrization is: 
\begin{equation}\label{ICC_ph}
 ICC_A = \frac{\omega^2}{\omega^2+\phi^2+\Tilde{\sigma}}.
\end{equation}
Figure \ref{ICCs} shows the difference between the empirical mean pairwise $ICC$ between each rater (red solid line) and the others and the computed $ICC_A$ (blue solid line) across independent datasets and different reliability scenarios. The mean difference between these two indices is consistently tight, and it seems to be narrower at increasing reliability levels. 

\paragraph{BNP ICC.}
The moments defined in (\ref{Moments_1}), (\ref{Moments_2}), and (\ref{Moments_3}) account for heterogeneous populations of subjects and raters and can be used to compute a flexible ICC. 
\begin{prop}\label{prop_1}
Given a random subject $i=1,\dots,I$, independently scored by two random raters $j,j' \in \mathcal{R}_i, j \neq j'$, the conditional correlation between the scores $Y_{ij}$ and $Y_{ij'}$ is: 
\begin{equation}\label{ICC_aj}
 ICC_{j,j'} = Corr\left(Y_{ij}, Y_{ij'} |G,H, \sigma_j^2, \sigma_{j'}^2 \right) =  \frac{\omega^2_G}{\sqrt{\omega^2_G +\phi^2_H +\sigma_j^2}\sqrt{\omega^2_G +\phi^2_H +\sigma_{j'}^2} }.
\end{equation}
\end{prop}
The proof is reported in Appendix \ref{A}. However, a more general index, unconditioned on specific raters' parameters, might be more useful in practice. For this reason, we propose a $ICC_A$ index for this BNP class of models. To this aim, the variance of subjects' true score $\omega^*_G$ and the variance of raters' systematic bias $\phi^2_H$ can be directly plugged into the ICC formula. Since we have heteroscedasticity across raters, we need to take the expectation of raters' residual variance $\mathbf{E}[\sigma^2|H]=\Tilde{\sigma}^*_H$. Similarly to the above parametric case, we reparametrize (\ref{eq:3b_bis}) with:
\begin{equation}\label{eq:3b_modBNP}
     1/\sigma^2_j|\gamma,\beta   \overset{\mathrm{ind}}{\sim}  Ga\left(1+\gamma_j,\frac{1+\gamma_j}{\beta_j}\right), \quad j=1,\dots,J,
\end{equation}
and define:
\begin{equation}\label{Var_res_BNP}
    \mathbf{E}[\sigma^2_j|G,H]=\mathbf{E}[\sigma^2_j|H] = \Tilde{\sigma}_{H} =\sum_{k \geq 1} \pi_{2k} \Tilde{\sigma}_k,
\end{equation}
where $\Tilde{\sigma}_k = (1+\gamma_k)/(\beta_k \gamma_k)$ is the mean residual variance for the $k$-th component of the infinite mixture. 
As a result, the $ICC_A$ for the BNP models might be computed as reported below. 
\begin{prop}\label{prop_2}
Given a random subject $i=1,\dots,I$, independently scored by two random raters $j,j' \in \mathcal{R}_i, \; j \neq j'$, satisfying $\sigma_j^2=\sigma_{j'}^2=\Tilde{\sigma}_H$: 
\begin{enumerate}[label=(\roman*)]
    \item \label{prop_2:1} the conditional correlation between the ratings $Y_{ij}$ and $Y_{ij'}$ is: 
\begin{equation}\label{ICC_a}
 ICC_A = Corr\left(Y_{ij}, Y_{ij'} |G,H, \sigma_j^2=\sigma_{j'}^2=\Tilde{\sigma}_H \right) =  \frac{\omega^2_G}{\omega^2_G +\phi^2_H +\Tilde{\sigma}_H}; 
\end{equation}
\item \label{prop_2:2} the $ICC_A$ is the lower bound of the conditional expectation of the correlation between the ratings $Y_{ij}$ and $Y_{ij'}$ (ICC):
\begin{equation}\label{ICC_LowBound}
 ICC_A \leq  \mathbf{E}[Corr\left(Y_{ij}, Y_{ij'}|G,H \right)] = \mathbf{E}[ICC|G,H]
\end{equation}
\end{enumerate}
\end{prop} 
The proofs are reported in Appendix \ref{A}. The index therefore accounts for the heterogeneity of the two populations (subjects and raters). It reduces to the parametric $ICC_A$ (\ref{ICC_p}) whenever $\omega^2_n=\omega^2$, for $n=1,\dots, \infty$; $\phi^2_k=\phi^2$ and $\Tilde{\sigma}_k=\Tilde{\sigma}$, for $k=1,\dots, \infty$; $ICC_A$ (\ref{ICC_a}) is a generalization of its parametric version (\ref{ICC_p}).  
The $ICC_A$ might reveal valuable information in inter-rater reliability or agreement analysis. For instance, when the ICC is used as an inter-rater reliability index \citep{Martinkova2023,ten2022updated, Erosheva2021}, the $ICC_A$ is the lower bound of the expected inter-rater reliability of a single rating. \\   
In this work, we mainly focus on the population level $ICC_A$, but different ICC indices can be computed and compared under this framework by conditioning on different subjects or raters' clusters. 


\section{Reduced Model for One-Way Designs}\label{Reduced}
 One-way designs are common when raters' identity is unknown and the systematic biases $\{\tau_j\}_1^J$ can not be identifiable. It might be seen as a limiting case in which each rater only scores one subject, i.e. $|\mathcal{S}_j|=1$. \\ 
Some blocks of the model in Section \ref{Main} reduce as briefly presented below. Note that we model subjects' true score $\theta_i$ as in the main model (\ref{eq:2}) and (\ref{DP1}).
 
 \paragraph{Modelling Rating $Y_{ij}$.} 
We decompose the observed rating $Y_i$ as: 
\begin{equation}\label{eq:1_reduced}
    Y_{ij} = \theta_i + \epsilon_{ij}, \quad i = 1,\dots,I; \;\; j \in \mathcal{R}_i, 
\end{equation}
Here $\epsilon_{ij}$ is the error of rater $j$ in rating the subject $i$ and it is the difference between the observed score $Y_{ij}$ and the subject true score $\theta_i$.  
 
\paragraph{Modelling Raters' error $\epsilon_n$}
 For each rating $Y_{ij}$ we assume that the rater's error $\epsilon_{ij}$ is drawn independently from a normal distribution with mean $\eta_{ij}$ and variance $\phi^2_{ij}$: 
 \begin{equation}\label{eq:error_reduced}
     \epsilon_{ij}|\eta_{ij},\phi^2_{ij} \overset{\mathrm{ind}}{\sim} N(\eta_{ij},\phi^2_{ij}), \quad i = 1,\dots,I; \;\; j \in \mathcal{R}_i.
 \end{equation}
We specify a DP prior with concentration parameter $\alpha_2$ and base measure $H_0$ for the two-dimensional vector $(\eta_{ij},\phi^2_{ij})$, for $i=1\dots,I$ and $ j \in \mathcal{R}_i$:
\begin{equation}\label{DP2_reduced}
    (\eta_{ij},\phi^2_{ij})|H \overset{\mathrm{iid}}{\sim} H, \quad \; H \sim DP(\alpha_2 H_0).
\end{equation}
We assume $\eta_{ij},\phi^2_{ij}$ to be independent and choose $H_0=N(\eta_0, D_0) \times IGa(a_0, A_0)$, where $\eta_0$ and $D_0$  are mean and scale parameters, respectively. This formulation induces a DPM prior for raters' error $\epsilon_{ij}$. 

\subsection{Identifiability and ICC}\label{Remarks_ICC_Reduced}
The moments of the error $\epsilon_{ij}$, $i=1\dots,I$ and $ j \in \mathcal{R}_i$,  are, respectively:
\begin{equation}\label{Moments_4}
 \mathbf{E}[\epsilon_{ij}|H] = \eta_{H} =\sum_{k\geq1} \pi_{2k} \eta_k, \quad \quad 
 \mathbf{Var}[\epsilon_{ij}|H] = \phi^2_{H} =\sum_{k\geq1} \pi_{2k} (\eta_k^2 + \phi_k^2) - \eta^2_{H}.  
\end{equation}
The centering strategy detailed in Section \ref{SC-DPM_sec} is here used and a SC-DPM is here placed over $\epsilon_{ij}$: 
\begin{equation}\label{SC-DPM_r}
   \epsilon^*_{ij} = \epsilon_{ij} - \eta_H,  \quad \quad 
      \epsilon_{ij}|\eta_{ij},\phi^2_{ij} \overset{\mathrm{ind}}{\sim} N(\eta_{ij},\phi^2_{ij}), \quad \quad i = 1,\dots,I; \;\; j \in \mathcal{R}_i.
\end{equation}
Under this parameter-expanded specification, the decomposition of rating $Y_{ij}$ (\ref{eq:1_reduced}) becomes: 
\begin{equation}\label{eq:1_exp_r}
     Y_{ij} = \theta_i +\epsilon^*_{ij}, \quad \quad \quad i = 1,\dots,I; \;\; j \in \mathcal{R}_i. 
\end{equation}
Given the location transformation in (\ref{SC-DPM_r}), the expectation of the residuals is zero: 
\begin{equation}\label{Exp_bias_new_r}
   \mathbf{E}[\epsilon^*_{ij}|H] = 0.
\end{equation}
For the one-way designs, the exact general ICC might be consistently estimated. \\
\begin{prop}\label{prop_3}
Given a random subject $i$, $i=1,\dots,I$, independently scored by two random raters $j, j' \in \mathcal{R}_i$, $j \neq j'$, the conditional correlation between the ratings $Y_{ij}$ and $Y_{ij'}$ is: 
\begin{equation}\label{ICC_reduced}
 Corr\left(Y_{ij}, Y_{ij'} |G,H \right) =  ICC = \frac{\omega^2_G}{\omega^2_G + \phi^2_H}.
\end{equation}
\end{prop}
The proof is given in Appendix \ref{A}. Conditioning on different clusters of subjects or raters and different ICC formulations lead to possible comparisons among clusters similar to the main model.

\section{Posterior Inference}\label{Inference}
The parameters of the DPs' base measures (i.e., $G_0$, $H_0$) and the respective concentration parameters $\alpha_1$ and $\alpha_2$ have to be assigned either a value or a hyperprior to complete the model specification and conduct posterior inference. This section outlines our choices about the hyperprior and the posterior computation. Several parameter specifications may be considered for the DP parameters \citep{Ghosal_van_der_Vaart_2017, Hjort_Holmes_Müller_Walker_2010} as they may be assigned a prior or fixed in advance. We placed a hyperprior on those parameters and let the data inform their parameters. \\
Under this model specification, the most natural choices to compute the posterior are conditional sampling schemes, such as Blocked Gibbs Sampling, which rely upon the approximate stick-breaking construction of the DP. They directly involve the prior in the sampling scheme avoiding its marginalization and accommodating hyperprior for the base measures \citep{Ishwaran2001}. They also come with further advantages, such as an improved mixing property, better interpretability of the mixture parameters \citep{Gelman2014, Hjort_Holmes_Müller_Walker_2010} and the direct computation of the ICC. Indeed, avoiding the prior marginalization, the moments (\ref{Moments_1}), (\ref{Moments_2}) and (\ref{Moments_3}) can be readily computed and plugged in the ICC formula (\ref{ICC_a}). \\
However, tailored considerations have to be made in practical applications based on specific data features.

\subsection{Hyperprior Specification.}
Eliciting the concentrations ' and base measures' parameters has a role in controlling the posterior distribution over clustering \citep{Gelman2014}. Small values of the variance parameters of the base measures $G_0$, and $H_0$ favor the clustering of subjects and raters, respectively, to different clusters. On the contrary, larger values of $G_0$ and $H_0$ variances favor the allocation of different subjects and raters, respectively, to the same cluster. \\  
We improve model flexibility by placing a prior on the base measures $G_0$ and $H_0$, and the concentration parameters $\alpha_1$ and $\alpha_2$ letting them be informed by the data.  
For the subjects' true score base measure $G_0=N(\mu_0, S_0) \times Ga(w_0, w_o/W_0)$ the following hyperpriors are specified:
\begin{equation}\label{prior_BaseMeasure_1}
    \mu_0 \sim N(\lambda_{\mu_0},\kappa^2_{\mu_0}), \quad 
    S_0 \sim IGa(q_{S_0},Q_{S_0}), \quad  w_0  \sim Ga(q_{w_0},Q_{w_0}), \quad W_0  \sim IGa(q_{W_0},Q_{W_0}). \nonumber
\end{equation}
We let $\lambda_{\mu_0}$ be the rating scale's center value (e.g., $\lambda_{\mu_0}=50$ on a 1-100 rating scale), $\kappa^2_{\mu_0}=100$ and the parameters $q_{w_0}, Q_{w_0},q_{W_0}, Q_{W_0}$ equal to 0.005. For the raters' base measure $H_0=N(\eta_0, D_0) \times Ga(a_0, a_0/A_0) \times Ga(b_0, b_0/B_0) \times Ga(m_0, m_0/M_0)$, the following hyperpriors are specified:
\begin{equation}\label{prior_BaseMeasure_2}
    \eta_0 \sim N(\lambda_{\eta_0},\kappa^2_{\eta_0}), \quad 
    D_0 \sim IGa(q_{D_0},Q_{D_0}), \quad  a_0  \sim Ga(q_{a_0},Q_{a_0}), \quad A_0  \sim IGa(q_{A_0},Q_{A_0}), \nonumber
\end{equation}
\begin{equation}\label{prior_BaseMeasure_2_bis}
     b_0  \sim Ga(q_{b_0},Q_{b_0}), \quad B_0 \sim IGa(q_{B_0},Q_{B_0}), \quad m_0 \sim Ga(q_{m_0},Q_{m_0}), \quad M_0 \sim IGa(q_{M_0},Q_{M_0}).
        \nonumber
\end{equation}
Where $\lambda_{\eta_0}=0$, $\kappa^2_{\eta_0}=100$, and the other hyperparameters are fixed to 0.005. \\
The concentration parameters $\alpha_1$ and $\alpha_2$ are assumed to follow respectively a gamma distribution: 
\begin{equation}\label{prior_concentrations}
    \alpha_1 \sim Ga(a_1,A_1) \quad \quad \alpha_2 \sim Ga(a_2, A_2).  \nonumber
\end{equation}
where $a_1,A_1,a_2,A_2$ are fixed to 1. The values we fix for the hyperprior's parameters are very common in literature and they are consistent with those proposed by many other studies on BNP models (e.g., \citealt{Mignemi2024, Paganin2023, Gelman2014,heinzl2012, Dunson2010}). 

\subsection{Posterior Computation.}
Since most of the parameters in the model have conjugate prior distributions, a Blocked Gibbs sampling algorithm was used for the posterior sampling \citep{Ishwaran2001}. No conjugate priors are available for the gamma's shape parameters (e.g., $\gamma_k$, $k=1,\dots, R$, $a_0$, $b_0$), thus we approximate the full conditionals using a derivatives-matching procedure (D-M) which is involved as an additional sampling step within the MCMC. This method has several advantages over other sampling schemes (e.g. adaptive rejection sampling or Metropolis-Hasting) in terms of efficiency, flexibility, and convergence property \citep{Miller2019}. We use the same D-M algorithm introduced by \citealt{Miller2019} to approximate the posterior of the gamma shape parameters of the base measures, i.e. $w_0, a_0, b_0, m_0$ and a modified version for the parameters $\gamma_k$, $k=1,\dots, R$, since the parametrization (\ref{eq:3b_modBNP}) is adopted.
We detail this adapted version of the D-M algorithm in the paragraph below and provide the complete Gibbs sampling in Supplementary Materials. 
\\
The notation on the independent allocation of subjects and rater to the corresponding clusters is  introduced here. Let $c_{1i}$ denote the cluster allocation of subject $i=1,\dots, I$, with $c_{1i}=n$ whenever $\xi_i=\xi_n$, $n=1,\dots, R$. Given the finite stick-breaking approximation detailed in Section \ref{finite_sb}, $R$ is the maximum number of clusters.  
We indicate the set of all the subjects assigned to the $n$-th cluster with $\mathcal{C}_{1n}$ and with $N_{1n}=|\mathcal{C}_{1n}|$ its cardinality.
Accordingly, let $c_{2j}$ denote the cluster allocation of rater $j=1,\dots, J$, such that $c_{2j}=k$ whenever $\zeta^*_j=\zeta^*_k$, $k=1,\dots, R$. The set of all the raters assigned to the $k$-th cluster is denoted by $\mathcal{C}_{2k}$ with $N_{2k}=|\mathcal{C}_{2k}|$ being its cardinality.

\paragraph{Derivatives-Matching Procedure.}\label{D-M}
Since no conjugate priors are available for the gamma's shape parameters $\{\gamma_k\}_1^R$, we involve, for each of these parameters, a D-M procedure to find a gamma distribution that approximates the full conditional distribution of these parameters, when their prior is also a gamma distribution \citep{Miller2019}. \\
We aim to approximate $p(\gamma_k|\cdot)$, i.e. the true full conditional density of $\gamma_k$, $k=1,\dots, R$, by finding $U_{1k}$ and $U_{2k}$ such that: 
\begin{equation}\label{deriv}
    p(\gamma_k|\cdot) \approx g(\gamma_k|U_{1k},U_{2k}), \quad k=1,\dots,R,
\end{equation}
where $g(\cdot)$ is a gamma density, $U_{1k}$ and $U_{2k}$ are shape and rate parameters, respectively. The algorithm aims to find  $U_{1k}$ and $U_{2k}$ such that the first and the second derivatives of the corresponding log densities of $p(\gamma_j|\cdot)$ and $g(\gamma_k|U_{1k}, U_{2k})$ match at a point $\gamma_{k}$. \citet{Miller2019} suggest to choose $\gamma_{k}$ to be near the mean of $p(\gamma_k|\cdot)$ for computational convenience. The approximation is iteratively refined by matching derivatives at the current $g(\cdot)$ mean as shown by Algorithm \ref{alg:cap}. We adapt the algorithm to our proposal, more specifically we consider the model involving the shape constraint introduced in equation (\ref{eq:3b_modBNP}). When this constraint is not imposed, the original algorithm by \citet{Miller2019} may be directly used. \\
We denote with $X_{1k}$ and $X_{2k}$ the sufficient statistics for $\gamma_k$ corresponding to the $k$-th raters' mixture component.  
For the implementation of the Algorithm \ref{alg:cap} we set the convergence tolerance $\epsilon_0=10^{-8}$ and the maximum number of iterations $M=10$.
Here $\psi(\cdot)$ and $\psi'(\cdot)$ are the digamma and trigamma functions, respectively.\\
The parameters $U_{1k}$ and $U_{2k}$, returned by the algorithm, are used to update $\gamma_k \sim Ga(U_{1k},U_{2k})$, $k=1,\dots,R$, through the MCMC sampling. The derivation of the algorithm is given in the Supplementary Materials. 

\begin{algorithm}
\caption{D-M Algorithm}\label{alg:cap}
\begin{algorithmic}
\State $X_{1k} \gets \sum_{j \in \mathcal{C}_{2k}} \log(1/\sigma^2_j)$ 
\State $X_{2k} \gets  \sum_{j \in \mathcal{C}_{2k}} 1/\sigma^2_j$
\State $T_k \gets X_{2k}/\beta_k - X_{1k} +N_{2k}\;\log(\beta_k)-N_{2k}$
\State $U_{1k} \gets b_0 + N_{2k}/2$
\State $U_{2k} \gets B_0 + T_k$

\For{$m=1,\dots,M$}
\State $\gamma_{k}=U_{1k}/U_{2k}$
\State $U_{1k} \gets b_0 + N_{2k}\;\gamma_{k}^2\;\psi'(1+\gamma_{k}) - N_{2k}\;\gamma_{k}^2/(1+\gamma_{k}) $
\State $U_{2k} \gets B_0 + (U_{1k}-b_0)/\gamma_{k} \;N_{2k}\;\log(1+\gamma_{k})+N_{2k}\;\psi(1+\gamma_{k})+T_k$
\If{$|\gamma_{k}/(U_{1k}/U_{2k})|<\epsilon_0$}
    \State \Return  $U_{1k}, U_{2k}$    
\EndIf
\EndFor
\end{algorithmic}
\end{algorithm}

\subsection{Post-processing Procedures}
\paragraph{Semi-Centered DPM Processes.}\label{Post-processing}
The sampling scheme detailed in the Supplementary Materials provides draws under the noncentered DPM model. However, as discussed in Section \ref{SC-DPM_sec}, it is not identifiable, and we need to post-process the MCMC samples to make inferences under the SC-DPM parameter-expanded model \cite{Dunson2010}. Since it is a semi-centered model that naturally constrains the raters' systematic bias $\{\tau_j\}_1^J$ to have zero mean, a few location transformations are needed. After computing $\eta_H$ according to \ref{Moments_2} for each iteration, the samples of $\mu_0,\mu_G,\{\theta_i\}_1^I$ and $\{\tau_j\}_1^J$ are computed:
\begin{eqnarray}\label{Post_centering}
    \mu_0^* &=& \mu_0 + \eta_H, \nonumber \\
    \mu_G^* & = & \mu_G + \eta_H, \nonumber \\
   \theta_i^* & = & \theta_i + \eta_H, \quad \quad \text{for} \;\; i=1,\dots,I; \nonumber \\
    \tau_j^* & = & \tau_j - \eta_H,  \quad \quad \text{for} \;\; j=1,\dots,J. \nonumber
\end{eqnarray}
The first three are due to the location transformation of $\tau_j$ and have to be considered for inference purposes under the SC-DPM model. 

\paragraph{Posterior Densities and Clusters Point Estimates.}
Each density equipped with a BNP prior might be monitored along the MCMC by a dense grid of equally spaced points \citep{Mignemi2024, Gelman2014, Dunson2010}. Each point of the grid is evaluated according to the mixture resulting from the finite stick-breaking approximation at each iteration. At the end of the MCMC, for each point of the grid posterior mean and credible interval might be computed, and as a by-product, the pointwise posterior distribution of the density might be represented. \\
The BNP model provides a posterior over the entire space of subjects' and raters' partitions, respectively. However, we can summarize these posteriors and determine the point estimates of these clustering structures by minimizing the respective variation of information (VI) loss functions. We refer to \cite{Wade&Ghahramani_2018} and \cite{Meila_2007} for further details on VI and point estimates of probabilistic clustering. \\
As for every parameter of the model, we use the posterior distribution of the subjects' specific parameters for inference purposes. Point estimates of the subjects' true scores $\{\theta_i\}_1^I$, such as the posterior mean (i.e., \textit{expected a posteriori}, EAP) or the \textit{maximum a posteriori} (i.e., MAP), might be used as official evaluations (i.e., final grades), and the posterior credible intervals as uncertainty quantification around those values. The $ICC_A$ index (\ref{ICC_a}) can be computed at each iteration of the MCMC to get its posterior distribution, which might be used for inference purposes.

\paragraph{Computational Details.}
In the present work, both for the simulations and the real data analysis, similarly to previous works (e.g., \citealt{Paganin2023,heinzl2012}), the number of iterations is fixed to 80,000 (with a thin factor of 60 due to memory constraints), discarding the first 20,000 as burn-in. We fix the maximum number of clusters to be $R=25$ respectively for subjects' and raters' DPM priors \citep{Gelman2014}.
The package \textit{mcclust.ext} \citep{mcclust.ext} is used for the point estimate of the clustering structures based on the VI loss functions. We graphically check out trace plots for convergence and use the package \textit{coda} for model diagnostics \citep{DeIorio_2023, CODA}. Convergence is also confirmed through multiple runs of the MCMC with different starting values\footnote{CPU configuration: 12th Gen Intel(R) Core(TM) i9 12900H.}.

\section{Simulation Study}\label{Simulation}
We perform a simulation study to compare the performance of the proposed models (BNP and a nested version) over the standard parametric one, highlighting the strength of our method. Concerning the \textit{individual-specific level}, the three models are evaluated on the accuracy of the estimates of the individual-specific parameters they provide (i.e., how close $\theta_i$, $\tau_j$, $\sigma^2_j$ are to the respective true values). Regarding the \textit{population level}, we compare the estimated population distribution of the subjects' and raters' features and evaluate the predictive performance of the three methods across different scenarios.

\paragraph{BP model.} The first model is the Bayesian parametric one (BP model), which can be considered a reduction of the BNP model in which all the subjects and the raters are allocated to the same cluster, respectively, such that $\mu_i=\mu$ and $\omega_i^2=\omega^2$, for $i=1,\dots, I$, and $\eta_j=\eta$, $\phi_i^2=\phi^2$, $\gamma_j=\gamma$ and $\beta_j=\beta$ for $j=1,\dots, J$. This model might be obtained by fixing the maximum number of clusters $R=1$. \paragraph{BSP model.} The second model is the Bayesian semi-parametric one (BSP model), in which the normality assumption is relaxed for the subjects' true score such that we model $\{\theta_i\}_i^I$ as detailed in Section \ref{Main}, but we model the raters' effects $\{\tau_j, 1/\sigma_j^2\}_1^J$ as in the parametric model (i.e., they are all assigned to the same cluster). This implies $R=1$ only for the rater-related DPM. Since in this model both $G$ and $H$ are degenerate on a mixture of only one component, we refer to the structural parameter as $\mu_G=\mu$, $\omega^2_G=\omega^2$ and $\phi^2_H=\phi^2$.  
\paragraph{BNP model.} The third model is the BNP model presented in Section  \ref{Main} in which the normality assumption is relaxed both for subjects and raters. Under this model, subjects and raters are allowed to be respectively assigned to different clusters. \\[0.4cm]
Three data-generative processes are set up with different clustering structures for subjects and raters. The densities of the subject's true score and the rater's effects are either unimodal, bimodal or multimodal. This allows us to assess the extent to which BNP priors might mitigate model misspecification and the BNP model reduces to the parametric one when the latter is properly specified; this setup is consistent with other works on BNP modeling in psychometrics \citep{Paganin2023}.\\
We keep some features of the generated data similar to the real data set analyzed in Section \ref{Applications} (e.g., sample size, rating scale, ratings per subject), they are also comparable with those of other works on rating models \citep{Bartos24, Martinkova2023}. Additional simulation results on small sample size applications of our proposal are presented in the Supplementary Materials.  

\subsection{Setting}
We generate subjects' ratings on a continuous scale, $Y_{ij} \in (1,100)$, the number of subjects $I=500$ and raters $J=100$ are fixed, whereas the number of ratings per subject and the true generative model vary across scenarios. 

\paragraph{Generative Scenarios.}
We manipulate the number of ratings per subject to be $|\mathcal{R}_i| \in \{2,4\}$ for $i=1,\dots,I$, since in many real contexts (e.g., education, peer review) it is common for the subjects to be rated only by two or few more independent raters \citep{Zupanc2018}. \\
Data are generated as specified by equations \ref{eq:1}, \ref{eq:3b}, and one of the schemes below, according to the three different scenarios:

\begin{description}
    \item[Unimodal:] Under this scenario, subjects' true score and raters' effects densities are unimodal:
     \begin{equation}\label{Uni}
         \theta_i \overset{\mathrm{iid}}{\sim} N(50,50), \quad \quad (\tau_j, 1/\sigma_j^2) \overset{\mathrm{iid}}{\sim} N(0,25) \;Ga(10,10/0.15), \nonumber
     \end{equation}
  for $i=1,\dots,I$ and $j=1,\dots,J$. This corresponds to the standard BP model in which subjects' true scores are assumed to be i.i.d across subjects and raters' effects are drawn jointly i.i.d. across raters. 

  \item[Bimodal:] In this scenario, both subjects' and raters' populations are composed, respectively, of two different clusters: 
     \begin{eqnarray}\label{Bimodal}
         \theta_i &\overset{\mathrm{iid}}{\sim}& 0.7 \cdot N(39,50) + 0.3 \cdot N(75.6,30) ,  \nonumber \\
         (\tau_j, 1/\sigma_j^2) &\overset{\mathrm{iid}}{\sim}& 0.5 \cdot  N(-5,10) \;Ga(10,10/0.1) + 0.5 \cdot  N(5,5) \;Ga(10,10/0.2), \nonumber
      \end{eqnarray}
    for $i=1,\dots,I$ and $j=1,\dots,J$.
\item[Multimodal:] Under this scenario, both subjects and raters are assigned respectively to three clusters:
    \begin{eqnarray}\label{Multimodal}
         \theta_i &\overset{\mathrm{iid}}{\sim}& 0.2 \cdot N(35,50) + 0.2 \cdot N(45,20)  + 0.6 \cdot N(56.6,20) ,  \nonumber \\
         (\tau_j, 1/\sigma_j^2) & \overset{\mathrm{iid}}{\sim}&  0.4 \cdot  SN(-5,3.162,-5) \;Ga(10,10/0.15)  \nonumber \\
         &+& 0.4 \cdot  N(0,10) \;Ga(10,10/0.10)  \nonumber \\
         &+& 0.2 \cdot  N(10,10) \;Ga(10,10/0.20), \nonumber
    \end{eqnarray}
    for $i=1,\dots,I$ and $j=1,\dots,J$. Here $SN(\xi,\omega,\alpha)$ stands for the skew-normal distribution with location, scale and slant parameters, $\xi$, $\omega$ and $\alpha$, respectively. 
    
\end{description}

These scenarios mimic three different levels of heterogeneity. From an interpretative point of view, in the first scenario, all the subjects' true scores are concentrated around the center of the rating scale, and the raters are quite homogeneous in their severity and reliability. The heterogeneity of the subjects and the raters is only at the individual level since they are not nested with clusters. Under the second scenario, we introduce heterogeneity at the population level as both subjects and raters are assigned to different clusters, respectively. Here, we mimic the case in which subjects are clustered within two different levels of true score (e.g., low vs. high proficiency level), and raters are either systematically slightly more lenient and reliable or more severe and less reliable. Under the third scenario, subjects and raters are assigned, respectively, to three poorly separated clusters. This results in a highly negatively skewed distribution for the subjects' true score and a multimodal distribution for the raters' systematic bias. Figures \ref{Densities_sim} and \ref{ppp_sim}, Figure \ref{Densities_sim_parametric} in Appendix \ref{Plots}, and Figures 1, 2 and 3 in the Supplementary Materials show the respective true densities and the empirical distributions of the generated ratings. \\
Ten independent data sets are generated under the six scenarios resulting from the $2 \times 3$ design, for each data set, the standard parametric (BP), the semi-parametric (BSP) and the nonparametric (BNP) models are fitted. 

\paragraph{Model recovery assessment.}
Parameter recovery performance is assessed through the Root Mean Square Error (RMSE) and the Mean Absolute Error (MAE) computed respectively as the root mean square difference and the mean absolute difference between the posterior mean and the true value of the parameters across data sets. For the subject and raters specific parameters, i.e. $\{\theta_i\}_1^I$, $\{\tau_j, 1/\sigma^2_j\}_1^J$, RMSE, and MAE are average both across individuals and data sets. \\ 
For the sake of comparison across different scenarios, we report the standardized version of both indices (S-RMSE, S-MAE) for the structural parameters. More precisely, those related to $\mu$ and $\mu_G$ are divided by the mean value of the rating scale, i.e. 50; those regarding $\omega^2$, $\omega^2_G$, $\phi^2$, $\phi^2$, $\Tilde{\sigma}$, $\Tilde{\sigma}_H$ and the $ICC_A$ are divided by their true value. \\
The models' performance in recovering the density distributions of individuals' specific parameters is evaluated through visual inspection. We give an example of how different densities might lead to very different conclusions on the data generative process \citep{paganin2024computational, BDA3, steinbakk2009posterior}. Specifically, we draw new replications from the respective posterior predictive distributions and compare these samples to the original data. If the models capture relevant aspects of the data, they should look similar, and replications should not deviate systematically from the data. We measure discrepancy in central asymmetry through the statistic $T_1(y,\mu_G)=|y_{.25}-\mu_G|-|y_{.75}-\mu_G|$, where $y_{.25}$ and $y_{,75}$ are the first and the third quartile, and in the left tail weight by the statistics $T_2(y)=min(y)$.

\subsection{Results}
\begin{figure}
    \centering
    \subfigure{\includegraphics[scale= 0.52]{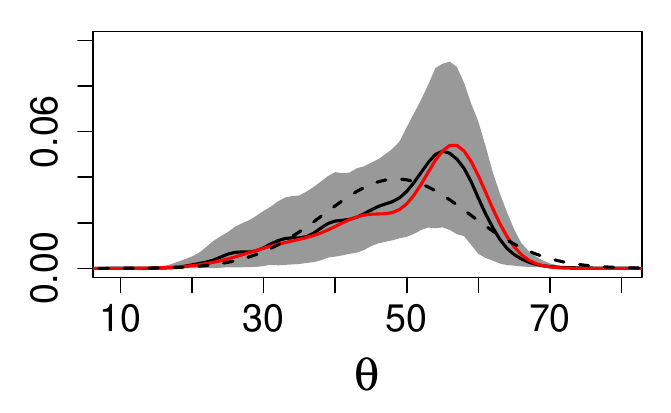}} 
    \subfigure{\includegraphics[scale= 0.52]{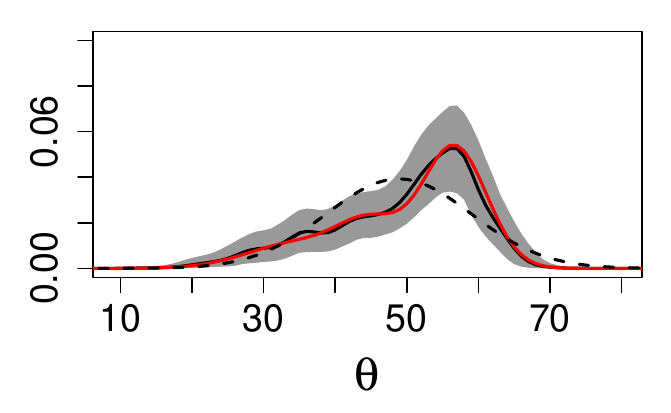}} 
     \subfigure{\includegraphics[scale=0.52]{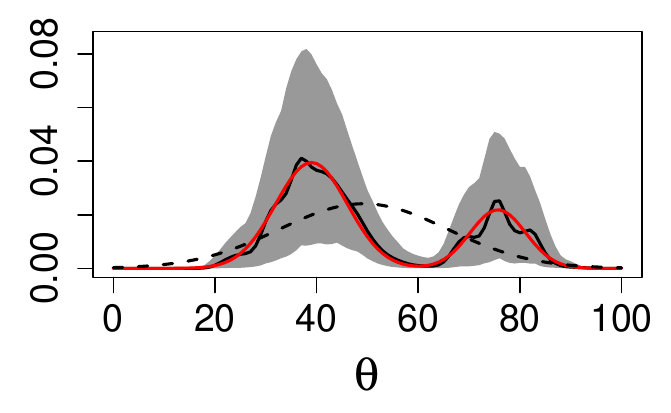}}
     \subfigure{\includegraphics[scale=0.52]{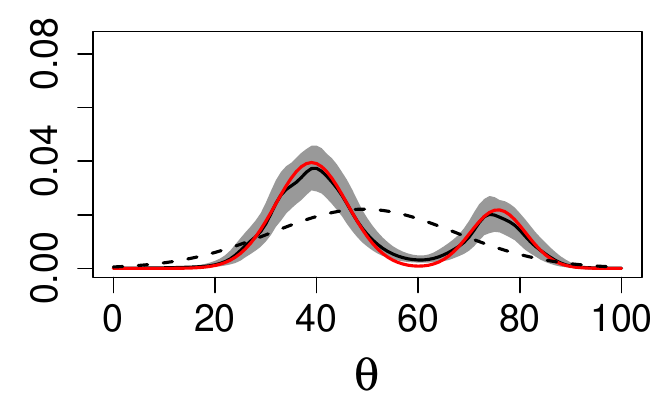}}
    \caption{\small Average estimated density across 10 independent datasets under different scenarios. The columns indicate the cardinality of $|\mathcal{R}_i|=\{2,4\}$: left and right, respectively; the rows indicate \textit{bimodal} or \textit{multimodal} scenario: first and second row, respectively. The solid red lines indicate the true densities; the solid black line and the shaded grey area indicate, respectively, the point-wise mean and $95\%$ quantile-based Credible Intervals; the density implied by the BP model (black dotted lines).  }
    \label{Densities_sim}
\end{figure}

\begin{figure}
    \centering
    \subfigure{\includegraphics[scale= 0.50]{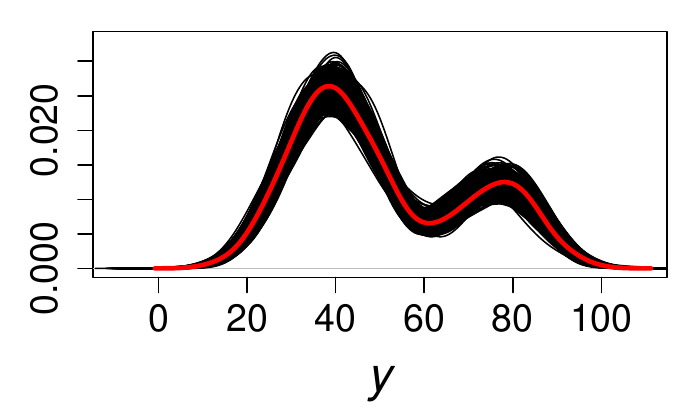}} 
    \subfigure{\includegraphics[scale= 0.50]{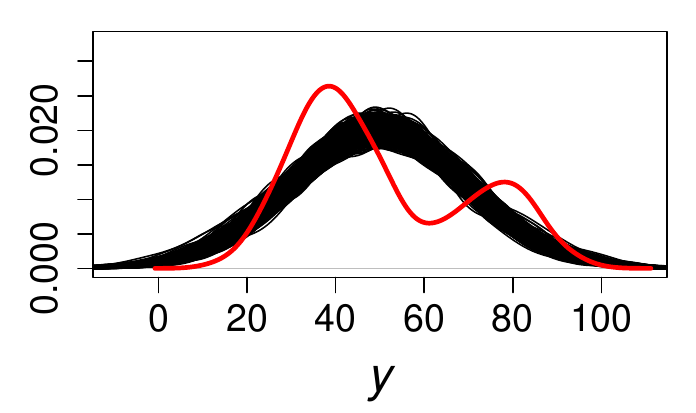}} 
    \subfigure{\includegraphics[scale= 0.50]{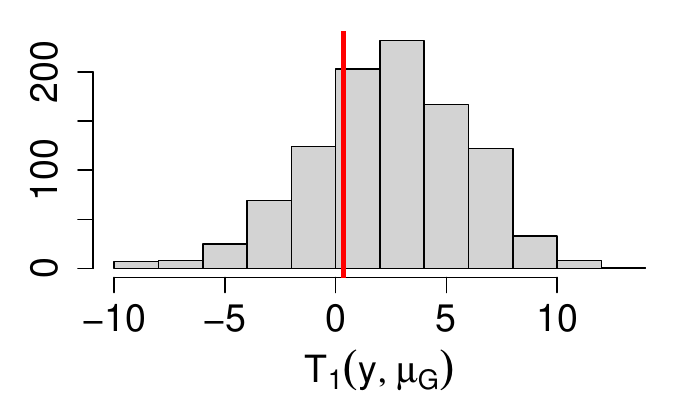}} 
    \subfigure{\includegraphics[scale= 0.50]{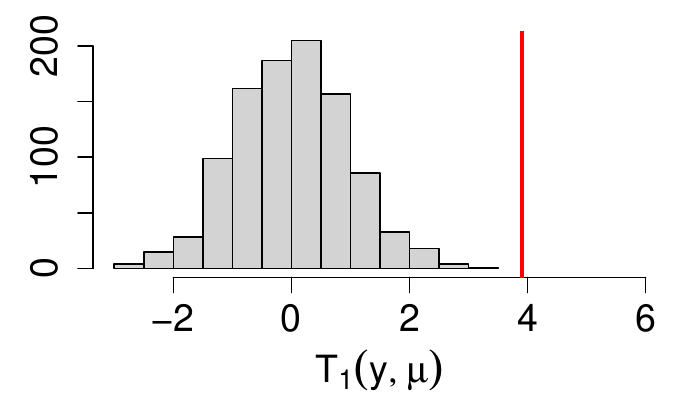}} 
     \subfigure{\includegraphics[scale=0.50]{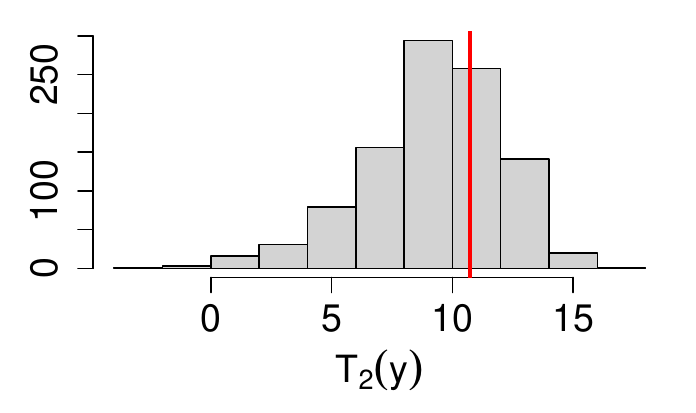}}
     \subfigure{\includegraphics[scale=0.50]{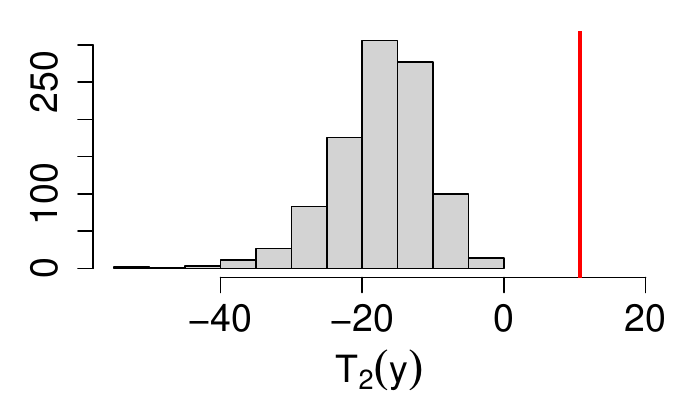}}
    \caption{\small Top row: empirical distribution of the data (red solid line) and empirical distribution of replicated data (black solid lines) from the respective BNP and BP posterior distributions (left and right columns, respectively).
    Middle and bottom row: Test statistics computed on the data (red solid line) and histograms of those computed on replicated data. }
    \label{ppp_sim}
\end{figure}

Results from the simulation study suggest that our proposals (i.e., BSP and BNP) systematically improve the estimates of the individual-specific parameters across scenarios. However, the accuracy of these estimates is comparable under the \textit{unimodal} scenarios across the three models. Meanwhile, the BSP and BNP models overcome, on average, the BP under the \textit{bimodal} and \textit{multimodal} scenarios in both conditions $|\mathcal{R}_i|=2$ and $|\mathcal{R}_i|=4$. As expected, the accuracy of subjects' and raters' specific parameters is higher in the conditions with a larger number of raters per subject $|\mathcal{R}_i| =4$ (Tables \ref{Tab:N2}). As indicated by $RMSE$ and $MAE$ indices, on average, the estimates of subjects' and raters' specific parameters provided by all the models degrade from the \textit{unimodal} to the \textit{multimodal} scenario. \\
Regarding the population parameter estimates, all the models provide overall similar estimates. We observe the largest improvement of the BSP and BNP over the parametric model under the \textit{bimodal} scenarios concerning subjects' true score variance $\omega_G^2$ and raters' systematic bias variance $\phi_H^2$. However, in these cases, the BP model provides better estimates of the expected residual variance $\tilde{\sigma}_H$. As a result, these differences are not detectable in the $ICC_A$ estimates and we observe equal accuracy for this index across the three models.\\
Figure \ref{Densities_sim} gives some examples of the estimated true score densities under the \textit{bimodal} and \textit{multimodal} scenarios; those under the \textit{unimodal} scenario are reported in the Appendix \ref{Plots}. The raters' features density plots are shown in the Supplementary Materials. The BNP model consistently estimates the respective densities under all the considered scenarios. The most prominent improvement of our proposals over the parametric model is observed under the heterogeneous scenarios. Accurate estimates of the densities are also provided under the extreme case of $|\mathcal{R}_i|=2$, that is, when each subject is rated by only two independent raters. Nonetheless, we note that the uncertainty about the densities is reduced when subjects are rated by a larger number of raters (i.e., $|\mathcal{R}_i|=4$). This reduction mostly regards the subjects' true score densities across all the scenarios. Our proposals capture the latent clustering structures of both subjects and raters as displayed by the posterior similarity matrices in Figure \ref{Post_sim_mat} in the Appendix \ref{Plots}. The entries of these matrices are the pairwise probability that two entries (e.g., subjects or raters) are clustered together. The clustering structure implied by the generative process under the \textit{bimodal} scenario is readily recognized by the graphical inspection. \\
The BNP model effectively captures relevant latent aspects of the data, such as deviations from normality both in the center and in the tails of the distributions across all the scenarios. As a by-product, the replications drawn from the posterior predictive distribution of the BNP model are remarkably more plausible than those generated under the BP model. As shown in Figure \ref{ppp_sim}, the normality assumptions made in the latter model restrict the shapes of the distributions for subjects' and raters' features. As a result, when these assumptions are violated, any inferences about the data-generating process might be misleading and unreliable. Replications under the BP model are far from the data both in the centre and on the tails of the distribution, as suggested by the statistics $T_1(y,\mu)$ and $T_2(y)$ in Figure \ref{ppp_sim}. \\
The improvement of our method over the parametric one is more prominent when the design is balanced (e.g., fully crossed designs) and the samples of subjects and raters are smaller. We present these results in the Supplementary Material.


\begin{table}
\centering
        \begin{tabular}{l c c c c c c c c}
           \toprule
          &  &   &\multicolumn{6}{c}{Generative Model} \\
            \midrule 
              &  & & \multicolumn{2}{c}{\textit{Unimodal}} &  \multicolumn{2}{c}{\textit{Bimodal}} & \multicolumn{2}{c}{\textit{Multimodal}}\\
           \midrule
                    &              &      &   RMSE          & MAE            &    RMSE & MAE &  RMSE & MAE \\ \midrule 
   
 $|\mathcal{R}_i| =2$ &   $\theta$   & BP  & \textbf{2.123} & 1.686          & 2.346          & 1.846          & 2.497          & 2.009 \\
                      &              & BSP & 2.127          & 1.689          & \textbf{2.308} & \textbf{1.822} & \textbf{2.347} & \textbf{1.889} \\
                       &             & BNP & \textbf{2.123} & \textbf{1.683} & 2.327          & 1.841          & 2.439          & 1.961 \\ \midrule 

                       & $\tau$      & BP  &  1.404          & 1.102          & 1.575          & 1.208          & 1.892          & 1.566 \\
                        &            & BSP &  1.407          & 1.104          & 1.554          & \textbf{1.206} & \textbf{1.700} & \textbf{1.387} \\
                         &           & BNP &  \textbf{1.401} & \textbf{1.101} & \textbf{1.553} & 1.212          & 1.774          & 1.460 \\ \midrule

                  & $1/\sigma^2$     & BP  & 0.070          & 0.060          & 0.092          & 0.076          & 0.085 & 0.070 \\
                          &          & BSP & \textbf{0.069} & \textbf{0.059} & \textbf{0.071} & 0.054          & \textbf{0.066} & \textbf{0.050} \\
                           &         & BNP & 0.071          &\textbf{0.059}  & \textbf{0.071} & \textbf{0.052} & \textbf{0.066} &  \textbf{0.050} \\ \midrule

$|\mathcal{R}_i| =4$ &  $\theta$   & BP    & 1.442          & 1.154          & 1.512          & 1.192          & 1.817          & 1.471 \\
                                  &  & BSP & 1.441          & 1.155          & 1.474          & 1.164          & 1.593          & 1.275 \\
                                 &   & BNP & \textbf{1.439} & \textbf{1.151} & \textbf{1.466} & \textbf{1.157} & \textbf{1.527} & \textbf{1.217} \\   \midrule

                       & $\tau$      & BP &  0.860          & 0.688          & 0.920          & 0.726           & 1.384          & 1.157 \\
                                &    & BSP & 0.860          & 0.686          & 0.886          & 0.711           & 1.088          & 0.885 \\
                               &     & BNP & \textbf{0.849} & \textbf{0.680} & \textbf{0.878} & \textbf{0.707}  & \textbf{0.996} & \textbf{0.798} \\\midrule

                  & $1/\sigma^2$     & BP  & 0.037 & 0.029 & 0.054          & 0.042          & \textbf{0.046} & 0.036 \\ 
                              &      & BSP & 0.037 & 0.029 & 0.054          & 0.041          & 0.048          & 0.037 \\ 
                               &     & BNP & 0.037 & 0.029 & \textbf{0.047} & \textbf{0.035} & 0.047          & \textbf{0.035} \\ 

 \bottomrule
        \end{tabular}
       
\caption[]{ \small Root Mean Square Error (RMSE) and Mean Absolute Error (S-MAE) of individuals parameters corresponding to Bayesian parametric model (BP), Bayesian semiparametric model (BSP) and Bayesian nonparametric model (BNP). }.
 \label{Tab:N2}
    \end{table}

\begin{table}
\centering
        \begin{tabular}{l c c c c c c c c}
           \toprule
        &    &  &\multicolumn{6}{c}{Generative Model} \\
            \midrule 
           &          & & \multicolumn{2}{c}{\textit{Unimodal}} &  \multicolumn{2}{c}{\textit{Bimodal}} & \multicolumn{2}{c}{\textit{Multimodal}}\\
           \midrule
            &          & & S-RMSE & S-MAE &  S-RMSE & S-MAE &  S-RMSE & S-MAE \\
            \midrule 
                 
 $|\mathcal{R}_i| =2$ &  $\mu$              & BP &   0.015  & 0.014          & 0.020           & 0.017          & 0.029          & 0.028  \\ 
                       & $\mu_G$            & BSP &  0.015  &\textbf{0.012}  & \textbf{0.014}  & \textbf{0.010} & \textbf{0.022} & \textbf{0.021}  \\ 
                       & $\mu_G$            & BNP &  0.015  & 0.013          & 0.016           & \textbf{0.010}  & 0.025         & 0.026  \\  \midrule
                    
                      & $\omega^2$         & BP  & \textbf{0.080} &  \textbf{0.066} & 0.284          & 0.284           & \textbf{0.064} & \textbf{0.052}  \\
                      & $\omega^2_G$       & BSP & 0.113          &  0.102          & \textbf{0.040} & \textbf{0.036}  & 0.080          & 0.072            \\
                      & $\omega^2_G$       & BNP & 0.094          &  0.080          & 0.065          & 0.045           & 0.110          & 0.094             \\ \midrule

                       & $\phi^2$           & BP &  0.979          &  0.110           &   2.343          & 2.341           & 0.273          & 0.239  \\ 
                       & $\phi^2$           & BSP & 0.152          &  0.111           &   0.161          & 0.142           & 0.261          & 0.229  \\ 
                       & $\phi^2_H$         & BNP & \textbf{0.134} &  \textbf{0.103}  &  \textbf{0.112}  & \textbf{0.084}  & \textbf{0.195} & \textbf{0.173}  \\ \midrule
                         
                       & $\Tilde{\sigma}$   & BP &  0.244          & 0.242          & \textbf{0.169} & \textbf{0.154} & 0.225          & 0.221 \\ 
                       & $\Tilde{\sigma}$   & BSP & 0.226          & 0.213          & 0.206          & 0.186          & \textbf{0.097} & \textbf{0.081} \\ 
                       & $\Tilde{\sigma}_H$ & BNP & \textbf{0.223} & \textbf{0.209} & 0.253          & 0.228          & 0.108          & 0.091 \\ \midrule

       &  $ICC_A$       & BP & 0.002 & 0.002 & 0.001 & 0.001 & 0.023 & 0.023 \\ 
        &             & BSP & 0.002 & 0.002 & 0.001 & 0.001 & 0.023 & 0.023 \\ 
        &             &  BNP & 0.002 & 0.002 & 0.001 & 0.001 &  0.023 & 0.23  \\  \midrule
 $|\mathcal{R}_i| =4$ &  $\mu$              & BP &   0.013          & 0.011  & 0.022          & 0.019          & 0.027            & 0.022  \\ 
                       & $\mu_G$            & BSP &  \textbf{0.012} & 0.011  & \textbf{0.017} & \textbf{0.014} & \textbf{0.018}   & \textbf{0.015}  \\ 
                       & $\mu_G$            & BNP &  \textbf{0.012} & 0.011  & 0.018          & \textbf{0.014} & 0.019            & \textbf{0.015}  \\  \midrule

                       & $\omega^2$    & BP   & \textbf{0.055} & \textbf{0.046} & 0.281          & 0.281          & \textbf{0.049} & \textbf{0.042} \\
                       & $\omega^2_G$  & BSP  & 0.108          & 0.092          & \textbf{0.046} & \textbf{0.043} & 0.066          & 0.052 \\
                       & $\omega^2_G$  & BNP  & 0.088          & 0.073          & 0.054          & 0.051          & 0.119          & 0.101 \\ \midrule

                       & $\phi^2$           & BP &  0.994          & 0.110          & 2.319          & 2.317          & 0.275          & 0.258 \\ 
                       & $\phi^2$           & BSP & 0.124          & 0.109          & 0.180          & 0.146          & 0.279          & 0.262 \\ 
                       & $\phi^2_H$         & BNP & \textbf{0.119}          & \textbf{0.105} & \textbf{0.132} & \textbf{0.095} & \textbf{0.209} & \textbf{0.188} \\ \midrule

                       & $\Tilde{\sigma}$   & BP &  \textbf{0.042}& \textbf{0.034} & \textbf{0.140} & 0.130          & \textbf{0.053} & \textbf{0.041} \\
                       & $\Tilde{\sigma}$   & BSP & 0.054         & 0.036          & 0.146          & 0.123          & 0.076          & 0.066 \\
                       & $\Tilde{\sigma}_H$ & BNP & 0.043         & 0.038          & 0.141          & \textbf{0.114} & 0.074          & 0.063 \\\midrule
         
       &  $ICC_A$       & BP & 0.002 & 0.002 & 0.001 & 0.001 & 0.023 & 0.023 \\ 
        &             & BSP & 0.002 & 0.002 & 0.001 & 0.001 & 0.023 & 0.023 \\ 
        &             &  BNP & 0.002 & 0.002 & 0.001 & 0.001 &  0.023 &  \textbf{0.022} \\
        
 \bottomrule
        \end{tabular}
      
\caption[]{ \small Standardized Root Mean Square Error (S-RMSE) and Standardized Mean Absolute Error (S-MAE) of structural parameters corresponding to Bayesian parametric model (BP), Bayesian semiparametric model (BSP) and Bayesian nonparametric model (BNP)}.
  \label{Tab:N4}
    \end{table}

\section{Application on large-scale essay assessment}\label{Applications}

\begin{figure}
    \centering
    \subfigure{\includegraphics[scale= 0.55]{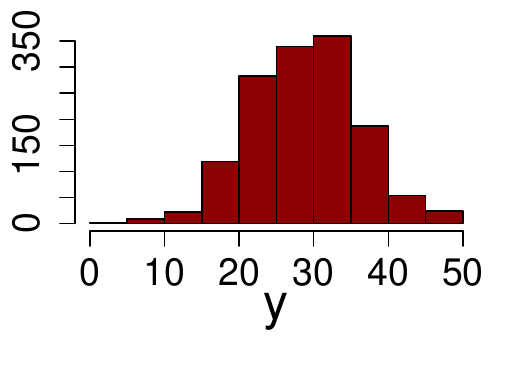}} 
     \subfigure{\includegraphics[scale=0.55]{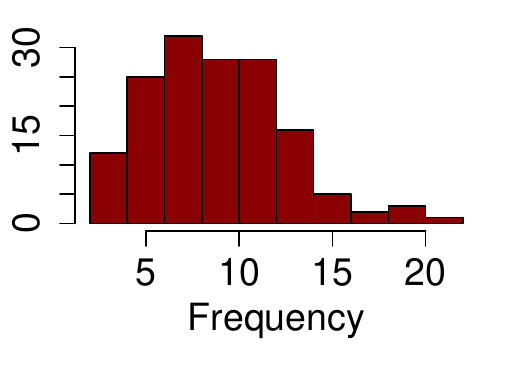}}
    \caption{\small The empirical distribution of ratings and the frequency of students per teacher are reported at left and right, respectively.}
    \label{Descriptives}
\end{figure}

We analyze the \textit{Matura} data set from \cite{Zupanc2018} as an illustrative example. 
The data come from a large-scale essay assessment conducted by the National Examination Centre in upper secondary schools in Slovenia during the nationwide external examination. 
Each student received a holistic grade on a 1-50 rating scale by two independent teachers.
We considered a random sample of $I=700$ students out of the 6995 who were examined during the spring term argumentative essays for the year 2014. A sample of $J=152$ teachers were involved who graded, on average, $9.21$ students, with a minimum of 2 and a maximum of 21 (see Figure \ref{Descriptives}). The observed ratings ranged from $0$ to $50$, with a mean of $29.35$, a skewness of $-0.051$, and a kurtosis of $3.148$ (see Figure \ref{Descriptives}). More details about the assessment procedure might be found in  \cite{Zupanc2018}.  \\

\paragraph{Model Comparison.} 
The three different models detailed in Section \ref{Simulation}, i.e. the parametric (BP) model, the semiparametric (BSP) model, and the nonparametric (BNP) model, were fitted to these data and compared on their out-of-sample prediction accuracy. The Watanabe–Akaike information criterion (WAIC) was used for this purpose. \\
This is a fully Bayesian approach for estimating the out-of-sample expectation, which relies on the computed log pointwise posterior predictive density and on a penalty term correction for the effective number of parameters to prevent overfitting \citep{Gelman2014}. The respective WAIC formulas are provided in the Supplementary Materials. 

\subsection{Results}
\begin{figure}
    \centering
    \subfigure{\includegraphics[scale= 0.51]{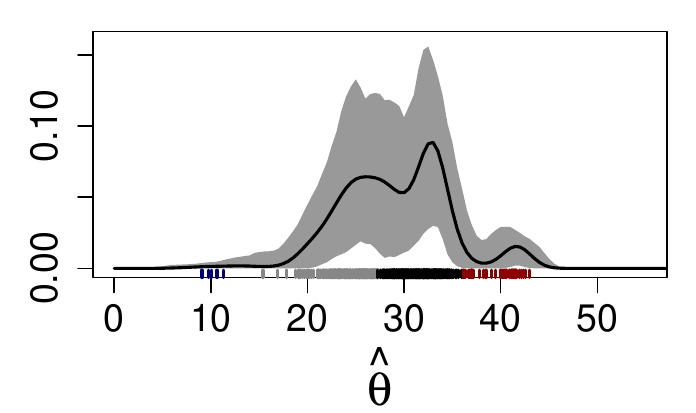}} 
    \subfigure{\includegraphics[scale= 0.51]{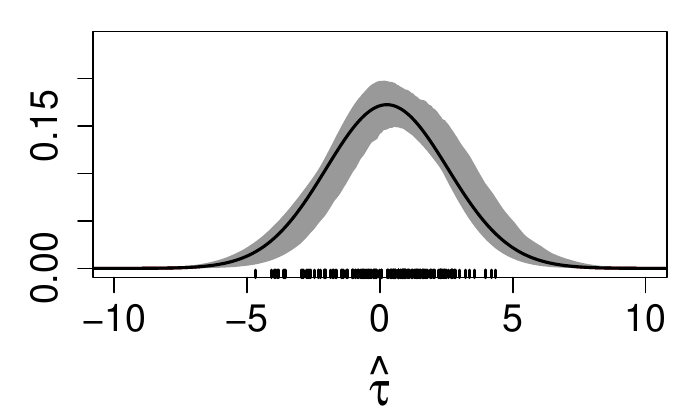}} 
     \subfigure{\includegraphics[scale=0.51]{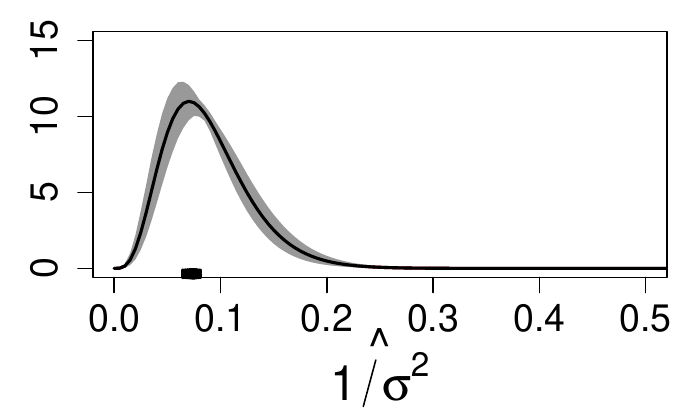}}
     \subfigure{\includegraphics[scale=0.51]{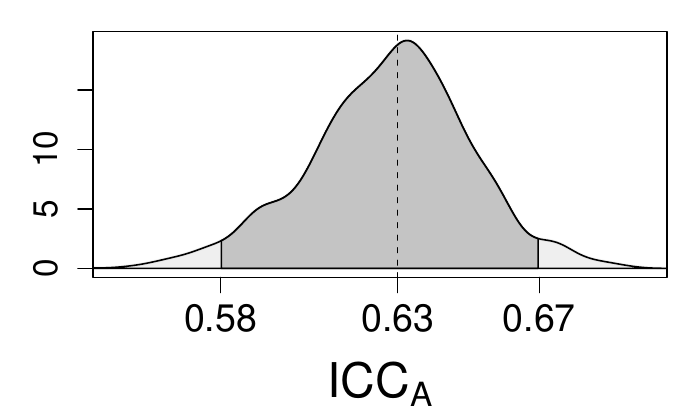}} 
    \caption{\small The estimated densities of the subject's true score $\theta$, rater's systematic bias $\tau$ and the residual term $\epsilon$ are reported; the black solid lines and the shade grey areas indicate the pointwise posterior mean and  $95\%$ quantile-based Credible Intervals of the respective densities. Bottom-right Figure shows the posterior distribution of the $ICC_A$, the black solid and dotted lines indicate, respectively, the $95\%$ credible interval and the posterior mean. The rugs at the margins of the first three Figures indicate the clustering of individuals.}
    \label{Densities_app}
\end{figure}

The total computational elapsed time for the BP, BSP, and BNP models was 180, 300, and 355 minutes, respectively. No convergence or mixing issues emerged from the graphical inspections of the MCMCs and diagnostics from \textit{CODA} package \citep{CODA}; further details and examples of trace plots are given in Supplementary Materials. Table \ref{Tab_WAIC} shows the WAIC indices for each fitted model and shows that the selection procedure indicates that the BNP model best fits the data and overcomes the others in predicting out-of-sample ratings. These results are consistent with the additional hold-out validation procedure presented in the Supplementary Materials. Based on the model comparison procedure, we focus on the results from the BNP model. \\
The posterior expectation of student ability mean $\mu_G$ and variance $\omega^2_G$ population parameters are $29.126$ and $32.702$, respectively. The respective narrow credible intervals suggest low uncertainty about these values. 
As expected from \citet{Antoniak_1974}, the posterior values of the concentration parameters $\alpha_1$ and $\alpha_2$ are proportional to the respective sample sizes and larger for the former. Details of the posterior values of base measures' parameters are reported in Supplementary Materials. 
The posterior expectation of raters' systematic bias variance $\phi^2_H$ and reliability $\Tilde{\sigma}_H$ are, respectively, $5.465$ and $13.913$. The corresponding credible intervals suggest low uncertainty around these values. \\
Figure \ref{Densities_app} gives the graphical representation of the respective estimated densities. The multimodal distribution of student ability $\theta$ implies heterogeneity among student abilities and points to the presence of multiple sub-populations. The variance in ratings is broadly due to students' ability, despite the variability of raters' systematic bias and reliability. Regarding the clustering structure of subjects and raters, the posterior similarity matrix, reported in Figure \ref{Post_sim_mat} in Appendix \ref{Plots}, suggests the presence of some latent partition of subjects, whereas no evidence of raters' clusters emerged from the posterior. This is coherent with the clusters' point estimate based on the variation of information (VI) loss function, which indicates four clusters for the subjects and one cluster for the raters. We render this result in Figure \ref{Densities_app} through rugs of different colors at the margin of the density plots; these values indicate the posterior mean of each subject and rater specific parameter. It is worth noting that we observe a cluster of subjects whose proficiency level is remarkably lower than the others, and another cluster in which subjects' performance is slightly superior than the others (Figure \ref{Densities_app}, upper-left; blue and brown rugs, respectively). These subjects might benefit from more personalized and specialized educational pathways.  
The posterior distribution of the $ICC_A$ with mean and credible intervals respectively equal to $0.627$ and $(0.577, 0.672)$, suggests a moderate inter-rater reliability; Figure \ref{Densities_app} shows the posterior distribution of this index. Since $ICC_A$ might be interpreted as the lower bound of the expected inter-rater reliability of a single rating, poor levels of reliability can be excluded \citep{koo2016}. However, this result is coherent with the findings of the original study by \cite{Zupanc2018}, where raters' variability and reliability have a substantial effect on ratings. Aggregate or average ratings over different teachers might mitigate inter-rater reliability issues \citep{Erosheva2021}.

\begin{table}
\centering
        \begin{tabular}{l c c }
           \toprule
          Fitted Model & $WAIC$ & $\Delta WAIC$ \\
            \midrule  
         BNP  Model    &     56267.43     &    -        \\
           BSP Model   &     67159.21      &   -10891.78    \\
            BP Model   &    168701.8      &    -112434.4       \\
            
 \bottomrule
        \end{tabular}
        \label{Tab_WAIC}
\caption[]{\small The Watanabe–Akaike information criterion (WAIC) is reported for each of the fitted models: Bayesian nonparametric model (BNP), Bayesian parametric (BP), and Bayesian semi-parametric (BSP); the pairwise WAIC difference ($\Delta WAIC$) between the model with the best fit and each other is reported.  }.
    \end{table}

\begin{table}
\centering
        \begin{tabular}{l l c c }
           \toprule
          &  & Posterior mean & $95\%$ Credible Interval\\
            \midrule 
      Subjects' parameters  & $\mu_G$              &  29.126   & $(27.886, 29.837)$  \\
                            & $\omega^2_G$           &  32.702   & $(28.198, 37.702)$  \\  
                         %
                         
                            &  $\alpha_1$          &  4.053   & $(0.915,9.218)$  \\ \midrule
        
      Raters' parameters    &   $\phi^2_H$           & 5.465   & $(4.085,7.476)$  \\ 
                            &   $\Tilde{\sigma}_H$ & 13.913    & $(12.424, 15.583)$ \\
                           
                         %
                         %
                            &  $\alpha_2$          & 1.839 & $(0.194,5.206)$  \\ \midrule

     & $ICC_A$ &    0.627 & $(0.577,0.672)$  \\ 
        
 \bottomrule
        \end{tabular}
        \label{Tab_2}
\caption[]{\small Posterior mean and $95\%$ quantile-based credible intervals of the estimated structural parameters of the BNP model are reported.}.
    \end{table}

\section{Coarsened Ratings Extension}\label{Ord_ex}
 Ratings data might be arbitrarily coarsened into a small number of ordered categories \citep{van2025estimation, CoarseGrades_2018, Coarse_2015a, Coarse_2015b}. As a result, continuous ratings that fall between two consecutive cut-offs are collapsed into the same ordered category, and fine-grained distinctions between individual scores are missing \citep{reardon2017_HETOP, Ho_coarsened}. The available ratings are ordinal in these cases, and the rating model proposed in Section \ref{Main} has to be modified accordingly.\\
 We leverage the underlying response variable formulation to extend the model to the ordinal case and consider the data coarsening mechanism \citep{Agresti2015, Nelson2015, Bartholomew2011, Cao_2010, Albert_Chib_93}. Our proposal might be seen as a BNP extension of the heteroscedastic ordered probit (HETOP; \citealt{Lockwood2018}). We specify the cumulative density function of the standard normal $\Phi(\cdot)$ as a link function, which implies that we only need to modify the equation (\ref{eq:1}). This extension might readily adapt to the One-Way designs presented in Section \ref{Reduced}. \\  
We note that coarse and ordinal ratings might be rather different. In the first case, the categories are consecutive intervals of a continuous rating scale, which is not the case for ordinal ratings. Here, we propose the HETOP specification as a possible straightforward extension of the main model for coarsened ratings and leave more advantageous formulations for ordinal data for future investigations.

\subsection{Categorical Modeling}
\paragraph{Modeling Rating $Y_{ij}$.}
We assume that the observed ordinal rating $Y_{ij}\in \{1,\dots, K\} \subset \mathbb{N}$ is generated by an underlying unobserved normally distributed variable $Y_{ij}^*$ \citep{joreskog_moustaki} and that we observe $Y_{ij}=k$ if $\delta_{k-1}<Y_{ij}^*\leq \delta_k$; $\delta_0=-\infty< \delta_1 < \dots, <\delta_K=+\infty$ are ordered thresholds over the underlying response variable distribution and are equal across raters. The underlying variable $Y_{ij}^*$ might be interpreted as a latent rating or the original continuous rating before the coarsening procedure. 
The conditional probability that $Y_{ij}=k$ is:
\begin{equation}\label{eq:1_ord}  
  \mathbb{P}[Y_{ij}=k|\theta_i, \tau_j,\sigma_j,\delta_{k},\delta_{k+1}] = \Phi \left( \frac{\delta_{k+1} - \theta_i - \tau_j}{ \sigma_j} \right) - \Phi \left( \frac{\delta_{k} - \theta_i - \tau_j}{ \sigma_j} \right),  
\end{equation}
for $\quad i = 1,\dots,I; \;\; j \in \mathcal{R}_i$. Additional considerations on the interpretation of $\sigma_j$ under this formulation are given in the Supplementary Materials. 

\paragraph{Identifiability issues.} 
Under this parametrization, we need to put additional constraints for identifiability purposes since the underlying response variables' mean and variance are freely estimated \citep{Kottas2018bayesian, kottas2005nonparametric}. Two thresholds (e.g., $\delta_1$, $\delta_{K-1}$ as proposed by \citealt{Song2013}) have to be fixed in advance, as it is common in multi-group analysis \citep{Lockwood2018}. From a statistical perspective, we note that each rater might be seen as a group of observations \citep{papaspiliopoulos2023}. Moreover, an SC-DPM prior has to be placed on the subject's true score $\{\theta_i\}_1^I$ to fix their mean and resolve identifiability issues \citep{Gelman2014}, as a by-product under the parameter-expanded specification, equation (\ref{eq:1_ord}) becomes: 
\begin{equation}\label{eq:1_ord_exp}  
  \mathbb{P}[Y_{ij}=k|\theta_i^*, \tau_j^*,\sigma_j,\delta_{k},\delta_{k+1}] = \Phi \left( \frac{\delta_{k+1} - \theta^*_i - \tau^*_j}{ \sigma_j} \right) - \Phi \left( \frac{\delta_{k} - \theta^*_i - \tau^*_j}{ \sigma_j} \right),
\end{equation}
for $\quad i = 1,\dots,I; \;\; j \in \mathcal{R}_i$. Whenever $K=2$, i.e. dichotomous rating scale, $\{\sigma_j\}_1^J$ can not be identified and need to be fixed in advance, e.g. $\sigma_j=1$, $j=1,\dots, J$, which implies assuming raters to be equally reliable \citep{Lockwood2018}. 

\paragraph{Generalized ICCs.}
Under this model specification, the ICCs computed according to propositions \ref{prop_1} and \ref{prop_2} are generalized intra-class correlation coefficients that indicate the polychoric correlation between two latent ratings \citep{joreskog1994_polychoric, Uebersax_1993}. For instance, proposition \ref{prop_1} implies here: 
\begin{equation}\label{ICC_aj_*}
 ICC_{j,j'}^* = Corr\left(Y_{ij}^*, Y_{ij'}^* |G,H, \sigma_j^2, \sigma_{j'}^2 \right) =  \frac{\omega^2_G}{\sqrt{\omega^2_G +\phi^2_H +\sigma_j^2}\sqrt{\omega^2_G +\phi^2_H +\sigma_{j'}^2} }
\end{equation}
where $ICC_{j,j'}^*$ indicates the conditional pairwise polychoric correlation between the latent ratings given by raters $j \neq j'$ to subject $i$. Similar considerations might be extended to propositions \ref{prop_2} and \ref{prop_3}. As a by-product, the $ICC_A^*$ is the lower bound of the expected polychoric correlation between the latent ratings $Y_{ij}^*$ and $Y_{ij'}^*$, with $j \neq j'$: 
\begin{equation}\label{ICC_LowBound_*}
 ICC_A^* \leq  \mathbf{E}[Corr\left(Y_{ij}^*, Y_{ij'}^*|G,H \right)] = \mathbf{E}[ICC^*|G,H].
\end{equation}

\subsection{Posterior computation}
A data augmentation procedure may simulate the underlying response variables \citep{Albert_Chib_93}. The underlying continuous ratings $Y_{ij}^*$, $i=1,\dots, I$, $j \in \mathcal{R}_i$ are sampled: 
\begin{eqnarray}
    Y^*_{ij}| \cdot & \overset{\mathrm{ind}}{\sim}  & N(\theta^*_i-\tau_j,\sigma^2_j)
   \times I(\delta_{k-1} < Y^*_{ij} \le \delta_k), \quad k=1,\dots,K. \nonumber
\end{eqnarray}
Here $I(\cdot)$ is an indicator function. Following \citet{Albert_Chib_93} the conditional posterior distribution of the $K-3$ freely estimated thresholds,e.g. $\delta_2,\dots,\delta_{K-2}$ might be seen to be uniform on the respective intervals: 
\begin{eqnarray}
   \delta_k|\cdot &\overset{\mathrm{ind}}{\sim}& U(max\{max\{Y_{ij}^*: Y_{ij}=k\}, \delta_{k-1}\}, min\{min\{Y_{ij}^*: Y_{ij}=k+1\}, \delta_{k+1}\} ), \nonumber
\end{eqnarray}
here $U(\cdot)$ stands for uniform distribution.\\
All the other parameters are updated according to the posterior sampling scheme detailed in Section 3.1 of Supplementary Materials and the post-process transformation outlined in Section \ref{Post-processing} needs to take into account the double-centering.  After computing $\mu_G$ and $\eta_H$ according to \ref{Moments_1} and \ref{Moments_2} for each iteration, the samples of $\mu_0,\mu_G,\{\theta_i\}_1^I$ and $\{\tau_j\}_1^J$ are computed as follows:
\begin{eqnarray}\label{Post_centering_ord}
    \mu_0^* &=& \mu_0 -\mu_G + \eta_H, \nonumber \\
   \theta_i^* & = & \theta_i - \mu_G + \eta_H, \quad \quad \text{for} \;\; i=1,\dots,I; \nonumber \\
    \tau_j^* & = & \tau_j - \eta_H + \mu_G, \quad \quad \text{for} \;\; j=1,\dots,J. \nonumber
\end{eqnarray}

\subsection{Generated and Real Coarsened Ratings Analysis}
In this Section we present the analysis of real and generated coarsened ratings and compare the results with those presented in Sections \ref{Simulation} and \ref{Applications}. For the real data, we deliberately coarsened the original continuous ratings analyzed in Section \ref{Applications} into $K=4$ ordered categories according to the following cutoffs: $\delta_1=20$, $\delta_2=30$, $\delta_3=40$. The fit of the BP, BSP and BNP models to the data are compared according to the WAIC for ordered data discussed in the Supplementary Materials. \\
We performed a simulation study to assess the accuracy of the BNP and the BP versions for ordered ratings. More specifically, the same data sets generated under the \textit{bimodal} scenarios in Section \ref{Simulation} are coarsened and considered for this study. We coarse these ratings into $K=4$ ordered categories according to three consecutive cutoffs: $\delta_1=35$, $\delta_2=50$, $\delta_3=75$. The same parameter recovery assessment procedure detailed in Section \ref{Simulation} is consistently used here. \\
In real context, the cutoffs of the coarsening procedure are generally known since the continuous rating scale is deliberately broken down into a small number of consecutive intervals and raters are explicitly asked to coarse their ratings accordingly \citep{van2025estimation, Coarse_2015b}. For example, on a 1-100 continuous scale, they might be asked to indicate which of the following intervals each subject's score falls into: (1-25), (25-50), (50,75) or (75,100). On the contrary, when ratings are directly given on an ordinal scale, the categories' labels are not necessarily associated with any continuous scale intervals (e.g., "poor", "acceptable", "good", "very good"). In these scenarios, we consider the observed ordered ratings as coarsened representations of underlying continuous values according to some unknown consecutive cutoffs. In the first case, this coarsening process is factual; in the second, it is merely assumed. However, since in both cases at least two cutoffs need to be fixed for identification purposes, we decide to fix $\delta_1$ and $\delta_3$ to the true values and let the model estimate $\delta_2$, both for real and generated data.   

\paragraph{Results.}
The total computational elapsed times for the BP, BSP, and BNP models were roughly similar to those of previous Sections. Upon graphical inspections of the MCMC chains and diagnostics, no convergence or mixing issues emerged for both generated and real data. Table \ref{Tab_WAIC_ord} gives the WAIC indices for each fitted model and suggests that the BNP model provides the best fit to the data. Based on this model comparison procedure, we focus on the results from the BNP model. As shown in Table \ref{Tab_2_ord}, the estimates are equivalent to those obtained under the continuous BNP model presented in Section \ref{Applications}. We note that the only notable difference concerns the point estimate of the subjects' clustering structure. In this case, they are clustered into two (instead of four) subjects' groups. \\
Results from generated data suggest that the BNP model provides more accurate estimates of subjects' and raters' specific parameters and overcomes the BP model. The only exception is observed for the rater-specific reliability parameter $1/\sigma^2$ under the scenario $|\mathcal{R}_i|=4$; here, the BP model overcomes our proposal. Under the standard parametric model, we only have two population parameters $\gamma$ and $\beta$ (i.e., $(\gamma_j,\beta_j)=(\gamma,\beta)$, for $j=1,\dots, J$) and, as a consequence, more information is available for their estimation. This might result in a faster accuracy improvement of this model for this set of parameters as the ratio of students per rater increases. The comparison between the $RMSE$ and the $MAE$ of Tables \ref{Tab:N2} and \ref{Tab_ordinal_sim} suggests that the estimates of both the BP and BNP models degrade with coarse data. The same trend emerged regarding the structural parameters and the densities; we report these results in the Supplementary Materials. 

\begin{table}
\centering
        \begin{tabular}{l l c c c c}
           \toprule
        &             & \multicolumn{2}{c}{$|\mathcal{R}_i| =2$}  & \multicolumn{2}{c}{$|\mathcal{R}_i| =4$} \\    \midrule  
        &            &   RMSE          & MAE  &   RMSE          & MAE        \\ \midrule 
   
  $\theta$    & BP  & 6.333          & 5.151          &  4.995          & 4.016           \\
              & BNP & \textbf{4.846} & \textbf{3.802} &  \textbf{3.677} & \textbf{2.883}  \\ \midrule 
 
 $\tau$       & BP  & 3.197          & 2.532          & 2.002          & 1.586 \\
              & BNP & \textbf{2.896} & \textbf{2.278} & \textbf{1.832} & \textbf{1.437} \\ \midrule 
 
 $1/\sigma^2$ & BP  & 0.195          & 0.184          & \textbf{0.065} & \textbf{0.054} \\
              & BNP & \textbf{0.104} & \textbf{0.080} & 0.097          & 0.074 \\
              \bottomrule
        \end{tabular}
   
\caption[]{ \small Root Mean Square Error (RMSE) and Mean Absolute Error (MAE) od individuals parameters across \textit{bimodal} scenarios with coarsened ratings.}.
 \label{Tab_ordinal_sim}
    \end{table}

\begin{table}
\centering
        \begin{tabular}{l c c }
           \toprule
          Fitted Model & $WAIC$ & $\Delta WAIC$ \\
            \midrule  
         BNP  Model    &  3798.11    &    -        \\
         BP Model      &  3815.65    &    -17.54       \\
         BSP Model     &  3897.22    &   -99.11    \\

 \bottomrule
        \end{tabular}
        \label{Tab_WAIC_ord}
\caption[]{\small The Watanabe–Akaike information criterion (WAIC) is reported for each of the fitted models: Bayesian nonparametric model (BNP), Bayesian parametric (BP), and Bayesian semi-parametric (BSP); the pairwise WAIC difference ($\Delta WAIC$) between the model with the best fit and each other is reported.  }.
    \end{table}

\begin{table}
\centering
        \begin{tabular}{l l c c }
           \toprule
          &  & Posterior mean & $95\%$ Credible Interval\\
            \midrule
                            & $\delta_2$           &  29.671    & $(28.932,30.291)$ \\
            \midrule
      Subjects' parameters  & $\mu_G$              &   29.678   & $(28.970, 30.384)$  \\
                            & $\omega^2_G$           & 30.513   & $(25.577, 36.228)$  \\  
                         %
                         
                            &  $\alpha_1$          &  4.174   & $(1.148,8.847)$  \\ \midrule
        
      Raters' parameters    &   $\phi^2_H$           & 5.958   & $(4.133, 9.395)$  \\ 
                            &   $\Tilde{\sigma}_H$ &  13.1080    & $(11.191, 15.351)$ \\
                           
                         %
                         %
                            &  $\alpha_2$          & 1.911 & $(0.237,5.249)$  \\ \midrule

     & $ICC_A$ &    0.627 & $(0.577,0.672)$  \\ 
        
 \bottomrule
        \end{tabular}
        \label{Tab_2_ord}
\caption[]{\small Posterior mean and $95\%$ quantile-based credible intervals of the estimated structural parameters of the BNP model are reported.}.
    \end{table}

\begin{figure}
    \centering
    \subfigure{\includegraphics[scale= 0.5]{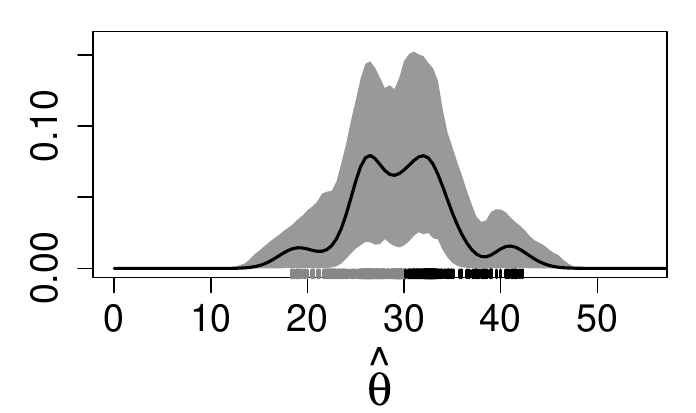}} 
    \subfigure{\includegraphics[scale= 0.5]{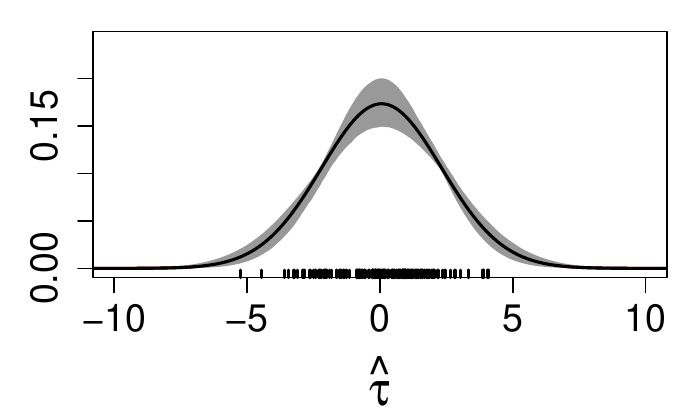}} 
     \subfigure{\includegraphics[scale=0.5]{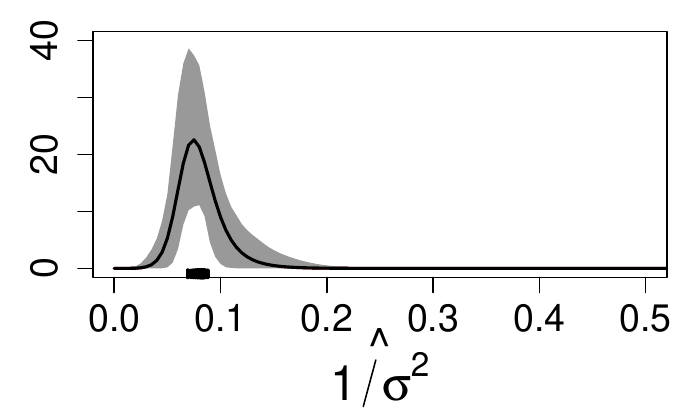}}
     \subfigure{\includegraphics[scale=0.5]{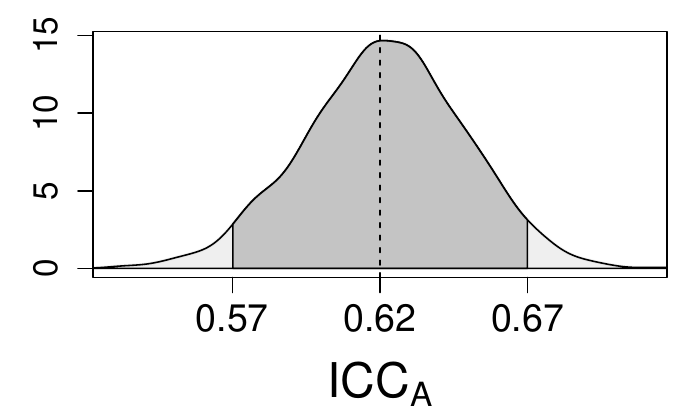}} 
    \caption{\small The estimated densities of the subject's true score $\theta$, rater's systematic bias $\tau$ and the residual term $\epsilon$ are reported; the black solid lines and the shade grey areas indicate the pointwise posterior mean and  $95\%$ quantile-based Credible Intervals of the respective densities. Bottom-right Figure shows the posterior distribution of the $ICC_A$, the black solid and dotted lines indicate, respectively, the $95\%$ credible interval and the posterior mean. The rugs at the margins of the first three Figures indicate the clustering of individuals. }
    \label{Densities_app_extension}
\end{figure}

\section{Concluding Remarks}\label{Discussion}
A flexible Bayesian nonparametric framework is proposed for the analysis of holistic rating data. We adopt the two-way unbalanced design as a general setting \citep{McGraw1996} which allows us to relate our proposal to other existing models (e.g., cross-classified or crossed random effects models, multilevel models, IRT-based rating models). We specify a measurement model to jointly estimate the subject's latent quality (e.g., student's proficiency) and the rater's features (i.e., severity and consistency). Our proposal may be suitable both for balanced (i.e. when all raters score each subject; \citealt{Nelson2015, Nelson2010}) and unbalanced designs (i.e. when a subset of raters scores each subject; \citealt{ten2022updated, Martinkova2023}). This method aims to capture latent heterogeneity among subjects and raters with the stochastic clustering induced by the Dirichlet Process Mixture (DPM) placed over their effects. This allows us to relax the common distributional assumptions on the respective parameters, preventing model misspecification issues \citep{antonelli2016, Walker2007}. \\
Results from the simulation study highlight the flexibility of our proposal, which provides accurate estimates across different scenarios. Exploiting the DPM prior, the respective densities of the students' and raters' effects are consistently estimated both when the normality assumption holds and when it is violated. Our method provides a more prominent improvement in small sample sizes and with coarse data. Our proposal provides the best fit to the real data, both for continuous and coarse ratings, compared to the parametric competitor. Nonetheless, the accuracy of the estimates with coarse ratings might be a concern when subjects are only rated by a very small number of raters and the estimated true scores are used, for instance, for selection purposes or as official grades. The theoretical results presented in Section \ref{ICC} are employed to make inferences about the inter-rater reliability of the single ratings.\\
The relatively long computational times of the MCMC chains might be prohibitive if used for repeated or massive scoring procedures. In such cases, if one is interested in capturing systematic heterogeneity among subjects or raters, any formulation of a mixture model (parametric or nonparametric) might be computationally cumbersome. In contrast, if this is not the focus of the analysis, the parametric model might be a computationally faster solution. \\
Under our model, rater's systematic bias and reliability are assumed to be independent conditional on the parameters of the cluster; additionally, the reliability of the raters is assumed to be independent of their specific workload $|\mathcal{S}_j|$ (i.e., the cardinality of the subset of subjects the rater has to evaluate). These assumptions might be unrealistic in some real contexts, and they might be relaxed under more general model specifications. For example, a multivariate distribution might be specified as a base measure $H_0$ to account for the correlation between the rater's features, and the rater-specific workload $|\mathcal{S}_j|$ might be modeled as a random variable correlated to the rater's features. Furthermore, because the measurement model includes raters' effects only as an additive component, all raters are assumed to have the same ability to discriminate between subjects with different latent true scores. This assumption might be relaxed by specifying an additional rater-specific multiplicative effect for the subject's true score, similar to the GMFRMs \citep{Uto_2016}. \\
The model detailed in Section \ref{Main} might be further extended to account for multidimensional ratings, i.e. when subjects are rated on multiple items. Under this three-way design, item parameters might be identified under some general conditions, and the model might extend \citealt{Paganin2023}, or \citealt{Karabatsos2009} to account for raters' characteristics. Further BNP generalizations of the existing rating models, e.g., GMFRMs, \citep[e.g.,][]{Uto_24, Uto_2016} HRMs \citep[e.g.,][]{Molenaar2021, Casabianca_2015,decarlo2011} or Trifactor Models \citep[e.g.,][]{soland2022, Shin_2019} are left for future investigations. \\
The effect of covariate and contextual factors might be incorporated in the structural models \ref{eq:2}, \ref{eq:3a}, or  \ref{eq:3b} if additional information on subjects or raters is available. This extension might relate our model to Explanatory Response Models \citep{Kim_2020,wilson2004} and be a BNP generalization of those methods. 
According to the data structure, more complex hierarchical priors might be placed over the subjects' true scores, such as hierarchical \citep{paisley2014,teh2004}, nested Dirichlet Process Mixtures \citep{rodriguez2008nested, Gelman2014, Hjort_Holmes_Müller_Walker_2010} or hidden hierarchical Dirichlet Process Mixtures recently introduced by \citet{Lijoi_2023} which overcomes some flaws of the previous ones. Stochastic Approximations of the DPM might be further considered for the stick-breaking constructions avoiding a maximum number of clusters \citep{Arbel2019}. \\
Our method might provide practitioners with valuable insights about the subjects' and raters' specific features along with the respective clustering structures. This information might be used to great advantage of individualized teaching programs \citep{coates_2025_precision} and might improve the matching procedure between subjects in peer teaching activities \citep{stigmar2016peer}.
Our theoretical finding and computational solution might enhance the analysis of rating data and contribute novel knowledge about the rating process.

\section*{\large  Acknowledgments }
We gratefully acknowledge the anonymous reviewers for their insightful suggestions regarding the extension to the coarse rating case. We are sincerely grateful to Professors Patricia Martinkova and Elena Erosheva for the precious insights on the first draft of the paper, and to Professors Antonio Lijoi and Igor Pruenster for fruitful discussions on this class of priors.

\begin{appendices}

\section{Further Extensions}
In this Section, we present some model extensions for more flexible clustering and complex hierarchical structures. We briefly detail alternative discrete priors that generalize the Dirichlet Process, and provide a more suitable framework for ratings collected across different populations of subjects or raters. 
\subsection{DP Generalizations}
Following the notation in Section \ref{Preliminaries}, given $\Pi=DP(\alpha P_0)$ the number of different unique values $K_n$ generated by $p$ increase asymptotically at a logarithmic rate, with $K_n \sim \alpha \; log(n)$ a.s. for $n \rightarrow \infty$. Alternative priors might be specified over $p$ which overcome this issue and allow for a more flexible prior specification on the number of clusters. More general specifications of $\Pi$ are briefly presented below. \\
Our proposal might readily encompass these priors, and since they all share the stick-breaking representation presented in Section \ref{StickBreaking1}, the $ICCs$ estimation and the Semi-centered identifiability procedure still hold for these cases. 

\paragraph{Mixture of Pitman-Yor Process.}
One of the most common generalizations of the Dirichlet Process is the Pitman-Yor Process $PY(d, \alpha, P_0)$, indexed by a discount parameter $0<d<1$, a concentration parameter $\alpha>-d$, and a base measure $P_0$. This is also termed the two-parameter Poisson Dirichlet process. For instance, we can place the $PY$ as a prior over the subject random measure $G \sim PY(d, \alpha, H_0)$, which might be represented as:
\begin{equation}\label{sum_PY}
   G = \sum_{n \geq1} \pi_{1n} \delta_{\xi_n}, \quad 
 \pi_{1n}     = V_{1n} \prod_{l<n}(1-V_{1l}),   \quad 
  V_{1n}    \overset{\mathrm{iid}}{\sim}    Beta(1-d,\alpha_1+nd),     \quad 
     \xi_n  \overset{\mathrm{iid}}{\sim} G_0, \nonumber 
 \end{equation}
Under this specification, the number of observed clusters $K_I$ out of a sample of $I$ subjects increase asymptotically at a rate $I^d$, with $K_I \sim S_{d,\alpha}I^d$ as $I \rightarrow \infty$. Here $S_{d,\alpha}$ is a limiting random variable with a probability distribution depending on $d$ and $\alpha$ and a positive density on $\mathbb{R}^+$. For $d \rightarrow0$, we recover the $DP(\alpha, G_0)$, whereas for larger values of $d$, the rate of increase of $K_I$ is faster. The discount parameter $d$ might be interpreted as the proportion of small clusters that will be observed out of a sample of $I$ subjects. Indeed, this parameter plays a double role in the clustering behavior of the model. The higher values of $d$ imply a \textit{reinforcement mechanism} that favors the allocation of a subject to the larger clusters (the `rich-get-richer' property) and, at the same time, a higher probability of being assigned to a new cluster. This is clear from $\mathbb{E}[\pi_{1n}]=O(n^{-1/d})$, for $0<d<1$, which suggests that the decay of the cluster sizes is governed by a power law. 

\paragraph{Mixture of Normalized Generalized Gamma Process.}
We can alternatively specify a Normalized Generalized Gamma (NGG) process as a prior for $p \sim NGG(\alpha, d, P_0)$ \citep{lijoi2007controlling, brix1999generalized}. This distribution is characterized by $\tau >0$, $d \in (0,1)$, and a base measure $P_0$. Following the previous example, we can consider the subject random measure to be distributed according to an NGG, $G \sim NGG(\alpha, d, G_0)$. It might be represented as:
\begin{equation}\label{sum_NGG}
   G = \sum_{n \geq1} \pi_{1n} \delta_{\xi_n}, \quad 
 \pi_{1n}  = T_n/\sum_{i \geq 1} T_i,   \quad 
     \xi_n  \overset{\mathrm{iid}}{\sim} G_0, \nonumber 
 \end{equation}
where $T_n$ are points of a generalized gamma process  with parameters $\alpha >0$, $d \in (0,1)$, and $\sum_{i \geq 1} T_i< \infty$ \citep{brix1999generalized}. For $d \rightarrow0$ we recover the Dirichlet Process. See \cite{Ghosal_van_der_Vaart_2017} for the correspondence between the PY and NGG processes. The interpretation of the parameters $\alpha$ and $d$, and the comments on the power law tails behavior of the PY process might be readily applied to the NGG process.\\
In educational rating contexts, the PY and the NGG processes might be preferred to the DP when the interest is to identify a few large clusters of subjects with similar proficiency levels and subjects who might need more \textit{one-on-one} or personalized teaching. We refer to \cite{DeBlasi2015, Hjort_Holmes_Müller_Walker_2010} and \cite{Ishwaran2001} for a broader treatment of this class of priors.

    \section{Proofs} \label{A}
\paragraph{Proof of Proposition \ref{prop_1}}
\begin{proof}
    Let $Y_{ij}$ and $Y_{ij'}$ be the ratings given by two random raters $j,j' \in \mathcal{R}_i, \;\; j \neq j'$, to a random subject $i$, for $i=1,\dots,I$: 
\begin{equation}
    Y_{ij} = \theta_i +\tau_j +\epsilon_{ij}, \quad \quad 
    Y_{ij'} = \theta_i +\tau_{j'} +\epsilon_{ij'}. \nonumber
\end{equation}
Assuming mutual independence between the terms of the decomposition: 
\begin{equation}
  \mathbf{Var}[ Y_{ij} |G,H] = \omega^2_G + \phi^2_H + \sigma_j^2, \quad \quad 
  \mathbf{Var}[ Y_{ij'} |G,H] = \omega^2_G + \phi^2_H + \sigma_{j'}^2 \nonumber
\end{equation}
and the conditional covariance between the two ratings is: 
\begin{eqnarray}\label{cond_covariance_1}
    \mathbf{Cov}[Y_{ij}, Y_{ij'} |G,H] & = & \mathbf{Cov}[\theta_i +\tau_j +\epsilon_{ij} , \theta_i +\tau_{j'} +\epsilon_{ij'} |G,H] \nonumber \\
    & = & \mathbf{Cov}[\theta_i,\theta_i|G,H] +  
    \mathbf{Cov}[\theta_i,\tau_{j'}|G,H] +  
    \mathbf{Cov}[\theta_i,\epsilon_{ij'}|G,H]+ \nonumber\\
  && \mathbf{Cov}[\tau_j,\theta_i|G,H]+ 
      \mathbf{Cov}[\tau_j,\tau_{j'}|G,H]+
       \mathbf{Cov}[\tau_j,\epsilon_{ij'}|G,H] + \nonumber\\
  &&   \mathbf{Cov}[\epsilon_{ij},\theta_i |G,H] + 
        \mathbf{Cov}[\epsilon_{ij},\tau_{j'} |G,H] +
        \mathbf{Cov}[\epsilon_{ij},\epsilon_{ij'}|G,H] \nonumber \\
    & = & \mathbf{Cov}[\theta_i,\theta_i|G,H] \nonumber\\
    & = & \omega^2_G. \nonumber
\end{eqnarray}
The correlation between the ratings is: 
\begin{eqnarray}\label{cond_correlation_1}
    ICC_{j,j'} = Cor[Y_{ij}, Y_{ij'} |G,H,\sigma_j^2,\sigma_{j'}^2] &=& \frac{ \mathbf{Cov}[Y_{ij}, Y_{ij'} ]|G,H}{\sqrt{ (\mathbf{Var}[ Y_{ij}|G,H ]} \sqrt{\mathbf{Var}[ Y_{ij'}|G,H ]) }} \nonumber \\
    &=& \frac{\omega^2_G}{\sqrt{\omega^2_G +\phi^2_H +\sigma_j^2}\sqrt{\omega^2_G +\phi^2_H +\sigma_{j'}^2} }.  \nonumber 
\end{eqnarray}
\end{proof}

\paragraph{Proof of statement \ref{prop_2:1} of Proposition \ref{prop_2}}
\begin{proof}
    Let $Y_{ij}$ and $Y_{ij'}$ be the ratings given by two random raters $j,j' \in \mathcal{R}_i, \;\; j \neq j'$, satisfying $\sigma_j^2=\sigma_{j'}^2=\Tilde{\sigma}_H$ to a random subject $i$, $i=1,\dots,I$: 
\begin{equation}
    Y_{ij} = \theta_i +\tau_j +\epsilon_{ij}, \quad \quad 
    Y_{ij'} = \theta_i +\tau_{j'} +\epsilon_{ij'}. \nonumber
\end{equation}
Assuming mutual independence between the terms of the decomposition: 
\begin{equation}
  \mathbf{Var}[ Y_{ij} |G,H] = \omega^2_G + \phi^2_H + \Tilde{\sigma}_H, \quad \quad 
  \mathbf{Var}[ Y_{ij'} |G,H] = \omega^2_G + \phi^2_H + \Tilde{\sigma}_H \nonumber
\end{equation}
and the conditional covariance between the two ratings is: 
\begin{eqnarray}\label{cond_covariance}
    \mathbf{Cov}[Y_{ij}, Y_{ij'} |G,H] & = & \mathbf{Cov}[\theta_i +\tau_j +\epsilon_{ij} , \theta_i +\tau_{j'} +\epsilon_{ij'} |G,H] \nonumber \\
    & = & \mathbf{Cov}[\theta_i,\theta_i|G,H] +  
    \mathbf{Cov}[\theta_i,\tau_{j'}|G,H] +  
    \mathbf{Cov}[\theta_i,\epsilon_{ij'}|G,H]+ \nonumber\\
  && \mathbf{Cov}[\tau_j,\theta_i|G,H]+ 
      \mathbf{Cov}[\tau_j,\tau_{j'}|G,H]+
       \mathbf{Cov}[\tau_j,\epsilon_{ij'}|G,H] + \nonumber\\
  &&   \mathbf{Cov}[\epsilon_{ij},\theta_i |G,H] + 
        \mathbf{Cov}[\epsilon_{ij},\tau_{j'} |G,H] +
        \mathbf{Cov}[\epsilon_{ij},\epsilon_{ij'}|G,H] \nonumber \\
    & = & \mathbf{Cov}[\theta_i,\theta_i|G,H] \nonumber\\
    & = & \omega^2_G. \nonumber
\end{eqnarray}
The correlation between the ratings is: 
\begin{eqnarray}\label{cond_correlation}
    ICC_A = \mathbf{Cor}[Y_{ij}, Y_{ij'} |G,H] &=& \frac{ \mathbf{Cov}[Y_{ij}, Y_{ij'} ]|G,H}{\sqrt{ (\mathbf{Var}[ Y_{ij}|G,H ]} \sqrt{\mathbf{Var}[ Y_{ij'}|G,H ]) }} \nonumber \\
    &=& \frac{\omega^2_G}{\omega^2_G + \phi^2_H + \Tilde{\sigma}_H}.  \nonumber 
\end{eqnarray}
\end{proof}
\paragraph{Proof of statement \ref{prop_2:2} of Proposition \ref{prop_2}}
\begin{proof}
Let us consider the function $ICC$ which, conditional on $G$ and $H$, is a convex function of the random variable $\sigma_j^2$: 
\begin{equation}
    ICC(\sigma^2_j|G,H) = \frac{\omega^2_G}{\omega^2_G + \phi^2_H + \sigma_j^2}, \quad j=1,\dots,J.
\end{equation}.
Let $ICC_A$ be the $ICC$ function of the expected value of $\sigma_j$:
\begin{equation}
    ICC_A = \frac{\omega^2_G}{\omega^2_G + \phi^2_H + \mathbf{E}[\sigma_j^2|G,H]}, \quad j=1,\dots,J.
\end{equation}
Note that $E[\sigma^2_j|G,H]=\mathbf{E}[\sigma^2_{j'}|G,H]$ for $j,j'=1,\dots,J$, $j\neq j'$. 
It readily follows from the conditional Jensen's Inequality that 
\begin{equation}
    ICC(\mathbf{E}[\sigma_j|G,H]) \leq \mathbf{E}[ICC(\sigma_j^2|G,H)].
\end{equation}
Since for brevity we define $ICC_A=ICC(\mathbf{E}[\sigma_j|G,H])$ and $ICC=ICC(\sigma_j^2|G,H)$:
\begin{equation}
    ICC_A \leq \mathbf{E}[ICC|G,H].
\end{equation}
where $\mathbf{E}[ICC|G, H]=\mathbf{E}[Corr\left(Y_{ij}, Y_{i,j'}|G, H\right)]$, $i=1,\dots, I$ and $j,j' in \mathcal{R}_i$. That is the expected correlation between two independent ratings given to a random subject. 
\end{proof}

\paragraph{Proof of Proposition \ref{prop_3}}
\begin{proof}
    Let $Y_{ij}$ and $Y_{ij'}$ be the ratings given by $j,j' \in \mathcal{R}_i, \;\; j \neq j'$, to a random subject $i$, $i=1,\dots,I$: 
\begin{equation}
    Y_{ij} = \theta_i +\epsilon_{ij}, \quad \quad 
    Y_{ij'} = \theta_i +\epsilon_{ij'}. \nonumber
\end{equation}
Assuming mutual independence between the terms of the decomposition: 
\begin{equation}
  \mathbf{Var}[ Y_{ij} |G,H] = \omega^2_G + \phi^2_H,  \quad \quad 
  \mathbf{Var}[ Y_{ij'} |G,H] = \omega^2_G + \phi^2_H  \nonumber
\end{equation}
and the conditional covariance between the two ratings is: 
\begin{eqnarray}\label{cond_covariance_2}
    \mathbf{Cov}[Y_{ij}, Y_{ij'} |G,H] & = & \mathbf{Cov}[\theta_i +\epsilon_{ij} , \theta_i +\epsilon_{ij'} |G,H] \nonumber \\
    & = & \mathbf{Cov}[\theta_i,\theta_i|G,H] + 
    \mathbf{Cov}[\theta_i,\epsilon_{ij'}|G,H]+ \mathbf{Cov}[\epsilon_{ij},\theta_i |G,H] + 
        \mathbf{Cov}[\epsilon_{ij},\epsilon_{ij'}|G,H] \nonumber \\
    & = & \mathbf{Cov}[\theta_i,\theta_i|G,H] \nonumber\\
    & = & \omega^2_G. \nonumber
\end{eqnarray}
The correlation between the ratings is: 
\begin{eqnarray}\label{cond_correlation_2}
    ICC = Cor[Y_{ij}, Y_{ij'} |G,H] &=& \frac{ \mathbf{Cov}[Y_{ij}, Y_{ij'} |G,H]}{\sqrt{ (\mathbf{Var}[ Y_{ij}|G,H ]} \sqrt{\mathbf{Var}[ Y_{ij'}|G,H ]) }} \nonumber \\
    &=& \frac{\omega^2_G}{\omega^2_G +\phi^2_H  }.  \nonumber 
\end{eqnarray}
Since the conditional variance of ratings is equal across subjects $\mathbf{Var}[ Y_{ij} |G,H]= \omega^2_G + \phi^2_H$ for $i=1,\dots, I$, the ICC is unique for all the subjects. 
\end{proof}

\section{Plots}\label{Plots}

\begin{figure}[H]
    \centering
    \subfigure{\includegraphics[scale=0.50]{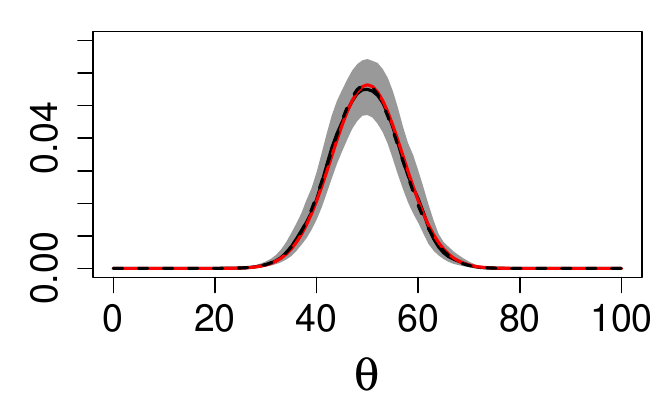}} 
    \subfigure{\includegraphics[scale=0.50]{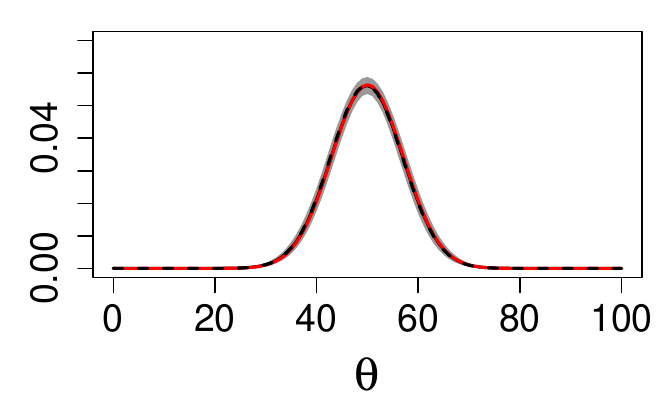}} 
    \caption{\small Average estimated density across 10 independent datasets under the \textit{unimodal} scenario. The columns indicate the cardinality of $|\mathcal{R}_i|=\{2,4\}$: left and right, respectively. The solid red lines indicate the true densities; the solid black line and the shaded grey area indicate, respectively, the point-wise mean and $95\%$ quantile-based Credible Intervals; the density implied by the BP model (black dotted lines). }
    \label{Densities_sim_parametric}
\end{figure}

\begin{figure}[H]
    \centering
    \subfigure{\includegraphics[scale= 0.60]{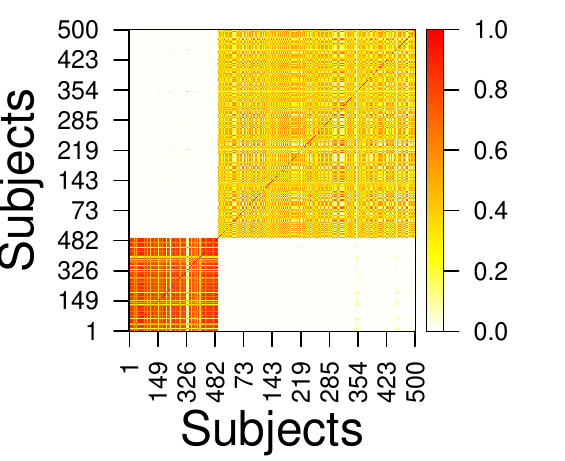}} 
    \subfigure{\includegraphics[scale= 0.60]{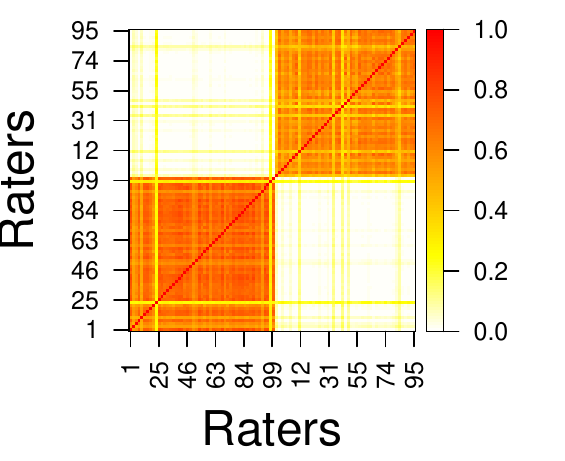}} 
     \subfigure{\includegraphics[scale=0.60]{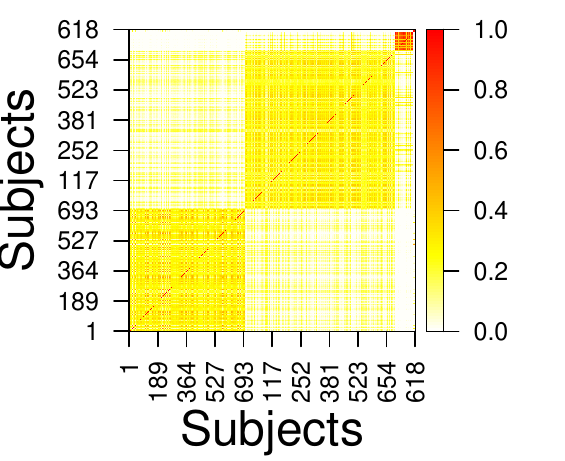}}
     \subfigure{\includegraphics[scale=0.60]{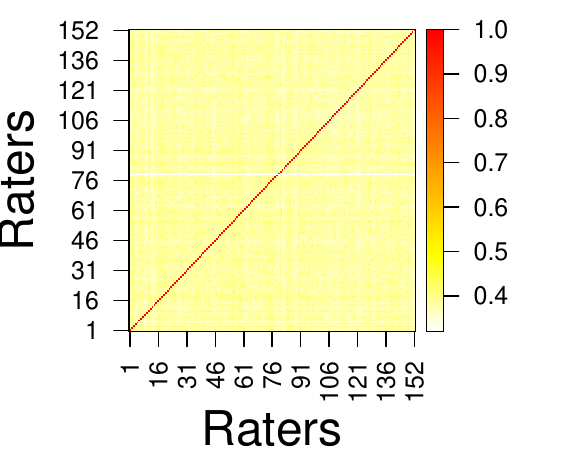}} 
    \caption{\small First row: examples of posterior similarity matrices for pairwise subject and raters allocation (left and right column, respectively). Second row: posterior similarity matrices for pairwise subject and raters allocation in real data analyzed in Section 7. }
    \label{Post_sim_mat}
\end{figure}

\end{appendices}

\bibliographystyle{apacite}

\bibliography{reference}

\end{document}